\documentclass[a4paper,fleqn]{cas-sc}

\usepackage[numbers]{natbib}
\usepackage{amssymb}
\usepackage{amsmath}
\usepackage{float}
\usepackage{xcolor}
\usepackage{siunitx}
\usepackage{tikz}
\usepackage{lineno}
\usepackage{subcaption}
\usepackage{pdfpages}
\usepackage{hyperref}
\usepackage{comment}

\def\tsc#1{\csdef{#1}{\textsc{\lowercase{#1}}\xspace}}
\tsc{WGM}
\tsc{QE}
\tsc{EP}
\tsc{PMS}
\tsc{BEC}
\tsc{DE}

\begin{document}
\let\WriteBookmarks\relax
\def\floatpagepagefraction{1}
\def\textpagefraction{.001}
\shorttitle{Precision Characterization for the SuperFGD scintillator cubes}
\shortauthors{A. Mefodiev et~al.}

\title [mode = title] {Precision Light Yield and Crosstalk Characterization for the SuperFGD scintillator cubes}

\author[a,b,e]{I.~Alekseev}
\author[b]{A.~Chvirova}
\author[a]{M.~Danilov}
\author[b]{S.~Fedotov}
\author[b]{A.~Khotjantsev}
\author[b,c]{M.~Kolupanova}
\author[d]{N.~Kozlenko}
\author[a]{A.~Krapiva}
\author[b,c,f]{Y.~Kudenko}

\author[a,b]{A.~Mefodiev}[orcid = 0000-0003-1243-0115]
\cormark[1]
\ead{mefodiev@inr.ru}

\author[b]{O.~Mineev}
\author[d]{D.~Novinsky}
\author[a,e]{E.~Samigullin}
\author[a,b,e]{N.~Skrobova}
\author[a,b,e]{D.~Svirida}

\affiliation[a]{organization={Lebedev Physical Institute of the Russian Academy of Sciences},
               addressline={53 Leninskiy Prospekt},
               city={Moscow},
               country={Russia}}

\affiliation[b]{organization={Institute for Nuclear Research of RAS},
               addressline={prospekt 60-letiya Oktyabrya 7a},
               city={Moscow},
               country={Russia}}

\affiliation[c]{organization={Moscow Institute of Physics and Technology},
               addressline={Institutskiy Pereulok, 9},
               city={Dolgoprudny},
               country={Russia}}

\affiliation[d]{organization={B.P. Konstantinov Petersburg Nuclear Physics Institute},
               addressline={Orlova Rocha, 1},
               city={Gatchina},
               country={Russia}}

\affiliation[e]{organization={National Research Center Kurchatov Institute},
               addressline={Ploshchad Akademika Kurchatova, 1},
               city={Moscow},
               country={Russia}}

\affiliation[f]{organization={National Research Nuclear University MEPhI},
               addressline={Kashira Hwy, 31},
               city={Moscow},
               country={Russia}}

\cortext[cor2]{Corresponding author}

\begin{abstract}
A detailed study of a $5\times5\times5$ cube prototype of the SuperFGD detector was performed using a 730~MeV/$c$ pion beam at the SC-1000 synchrocyclotron (PNPI, Gatchina, Russia). The detector, based on plastic scintillation cubes with orthogonal wavelength-shifting (WLS) fiber readout and silicon photomultipliers (SiPMs), was tested to evaluate its performance in terms of light yield, spatial uniformity, and optical crosstalk. Using high-resolution tracking, the spatial distribution of light yield was mapped with a granularity of 0.5~mm. An average light response map was obtained by combining data from 27 cubes. Optical crosstalk between adjacent cubes was also measured and characterized in four directions (left, right, up, down). Position-dependent crosstalk values ranged from 2\% to 6\%, with the highest levels observed near cube interfaces. These results confirm the excellent performance and scalability of the SuperFGD design, and provide valuable input for simulation tuning and reconstruction algorithms in the ND280 upgrade of the T2K experiment. The obtained result on the response uniformity and crosstalk are reasonably well described by simple MC model of the setup.
\end{abstract}






\maketitle
\setcounter{footnote}{0}

\newpage
\section{Introduction }
Scintillator-based detectors with high spatial granularity are widely employed in modern experiments due to their excellent tracking capabilities, robustness, and compatibility with magnetic fields. In neutrino physics, notable examples include the T2K Fine-Grained Detector~\cite{Amaudruz_2012}, the \mbox{MINERvA} detector~\cite{minerva}, the magnetized steel and scintillator calorimeters of the MINOS experiment~\cite{minos}, and the SoLid~\cite{solid} and DANSS~\cite{Alekseev:2016llm} detectors.

Different photodetector technologies have been employed to achieve the high granularity needed to reconstruct interaction vertices or to apply Particle Flow Algorithms (PFA). For example, a high-granularity plastic scintillator tile hadronic calorimeter with Avalanche Photo-Diode readout was proposed for a linear collider detector~\cite{apd_calice}. The invention of Silicon Photomultipliers~\cite{bondarenko,dolgoshein} made a revolution in this field. Detectors with wavelength-shifting (WLS) fiber light collection and SiPM readout, as well as detectors with direct SiPM readout developed by the CALICE collaboration~\cite{ahcal}, immediately demonstrated the great potential of this approach, although they initially used first-generation SiPMs. Since then, SiPM technology has matured and is now used in many fields. For example, the CMS collaboration replaced the photodetectors in the hadronic calorimeter with SiPMs, and a high-granularity scintillator tile detector with SiPM readout will be used in the high-luminosity upgrade of the CMS hadron calorimeter~\cite{cms_calice}, building on the approach initially developed by the CALICE collaboration for linear collider detectors~\cite{calice}. This technology allows a dramatic improvement in jet energy resolution through the PFA method. Furthermore, SiPMs provide excellent time resolution and are well suited for time-of-flight measurements~\cite{betancourt,korzenev,Alekseev:2022jki}.

Among recent developments, three-dimensional granular scintillator concepts have emerged as a powerful approach for imaging charged-particle tracks, electromagnetic and hadronic showers, and neutrino interactions. The SuperFGD~\cite{Kudenko:2025dlg,Dergacheva:2025zwe} represents a realization of this concept at scale, providing isotropic 3D tracking in the ND280 near detector of the T2K experiment. The detector consists of approximately two million $1\times1\times1$~cm$^3$ plastic scintillator cubes, with each cube viewed by three WLS fibers along orthogonal $X$, $Y$, and $Z$ directions. Each fiber is optically coupled to a silicon photomultiplier (SiPM), providing three-dimensional readout granularity.

The SuperFGD is the central active target of the upgraded ND280 near detector~\cite{nd280upgrade} in the T2K experiment, designed to improve the sensitivity of neutrino oscillation measurements. The former ND280 detector had limited angular acceptance, reconstructing only forward-going tracks, and a relatively high detection threshold for short-range particles  such as low-momentum protons. As a result, the hadronic final state of neutrino interactions was only partially reconstructed, and the neutrino energy was estimated using the muon momentum alone.

The SuperFGD overcomes these limitations through its fully isotropic three-dimensional tracking capability and fine granularity. It enables reconstruction of tracks at any angle, including backward-going particles, and significantly lowers the detection threshold for protons, allowing precise characterization of the hadronic vertex in charged-current quasi-elastic (CCQE) interactions. Furthermore, the upgraded ND280 provides neutron detection in anti-$\nu_\mu$ interactions, electron/gamma separation, and improved particle identification via time-of-flight measurements. Together, these improvements are expected to reduce the total systematic uncertainty for appearance modes from $\sim$6\% to $\leq$4\%, with an average reduction of $\sim$30\% on individual systematic parameters~\cite{nd280upgrade}.

A key feature of the SuperFGD is its ability to perform precise track reconstruction and calorimetric measurements in a highly segmented volume~\cite{Chvirova:2025vnw}. However, to fully exploit its potential, detailed knowledge of the light collection properties at the cube level is required. In particular, the spatial variation of the light yield within a cube, and optical crosstalk between adjacent cubes must be characterized and understood. These factors affect the accuracy of track reconstruction, the achievable energy and time resolution, and reconstruction of neutrino events.

In this work, we present results from a dedicated beam test of a small-scale prototype of the SuperFGD detector. A $5\times5\times5$ cube array, with 75 WLS fibers and 27 instrumented readout channels, was tested with a beam of 730~MeV/$c$ $\pi^{-}$ at the SC-1000 proton synchrocyclotron in PNPI (Gatchina, Russia). External proportional chambers were used to provide high-precision tracking of beam particles, allowing the reconstruction of the particle entry point into the detector with sub-millimeter accuracy.

The main objectives of this study are:
\begin{itemize}
\item To measure the light yield variation within a cube;
\item To evaluate the cube-to-cube uniformity;
\item To study the optical crosstalk between neighboring cubes;
\item To develop a Monte Carlo (MC) simulation of the light collection within a cube.
\end{itemize}

The results presented here are essential for the validation and tuning of the SuperFGD simulation, for the design of reconstruction algorithms, and for understanding of systematic effects in future data analyses.
\label{sec:introduction}
\section{SuperFGD cubes}
The $1\times1\times1\text{~cm}^3$ scintillator cubes were injection molded from polystyrene with addition of 1.5\% paraterphenyl (PTP) and 0.01\% POPOP at the UNIPLAST Co. (Vladimir, Russia)~\cite{uniplast}. After fabrication, each cube was coated with a highly reflective layer formed by etching the scintillator surface with a chemical agent. The etching process produced a white microporous polystyrene deposit on the surface, serving as a diffuse reflector. The thickness of the reflective layer was in the range of 50–80 $\mu$m. Three orthogonal through-holes of 1.5 mm diameter were drilled in each cube to accommodate WLS fibers.
\label{sec:cubes}
\section{Previous beam tests}
To validate the performance of the SuperFGD detector concept, several prototypes were tested with charged particles and neutrons. A compact $5\times5\times5$ array~\cite{Mineev:2019} and a larger $8\times24\times48$ array~\cite{Blondel_2020} were exposed to charged beams at the CERN PS T9 beamline. Another prototype of $8\times8\times32$ cubes was tested with neutrons at LANL~\cite{Agarwal:2022kiv}.

The CERN tests aimed to measure light yield (LY), timing resolution, and particle identification via $dE/dx$~\cite{Blondel_2020}. The LY varied between 36–50~p.e./cube/fiber, with a typical value of 80~p.e./cube/two fibers. Proton signals reached up to 450~p.e./cube/fiber. The average optical crosstalk was measured to be 3.7\%~\cite{Artikov2022}. The timing resolution was 0.97~ns for a single fiber and improved to 0.68~ns for a cube readout with two fibers~\cite{Alekseev:2022jki}.

At LANL, the SuperFGD prototype was used to detect neutrons up to 800~MeV via time-of-flight. These measurements enabled the first cross-section determination of neutrons on CH (hydrocarbon scintillator) over a broad energy range, demonstrating the detector’s relevance for neutrino energy reconstruction in T2K~\cite{PhysRevD.107.032012}.

\label{sec:Beam_tests}
\section{Experimental setup}
The beam test prototype was assembled as a $5\times5\times5$ array of the cubes covered by an outer protective shell made of thin, hard material. A total of 75 WLS fibers were installed through the holes of the cubes. The fibers were 1-mm diameter Y11(200) Kuraray S-type, each 35 cm in length. One end of each fiber was coupled to a SiPM with an optical custom made connector, another end was left open. The photosensors used were Hamamatsu MPPCs 12571-025C (table~\ref{tab:SiPM}).

\begin{table}[htbp]
\centering
\caption{\label{tab:SiPM} Main parameters for the MPPC type installed on the SuperFGD prototype.}
\smallskip
\begin{tabular}{|l|c|}
\hline
Description&  S12571-025C\\
\hline
Pixel pitch [$\mu m$] &  25\\
Number of pixels &  1600\\
Active area [$mm^2$] & $1\times1$ \\ 
Operating voltage [V] & $67-68$\\
Photon detection eff. [\%] & 35\\
Dark count rate [kHz] & 100\\
Gain & $5.15\times10^5$ \\
\hline
\end{tabular}
\end{table}

The signals from the SiPMs were amplified by a preamplifier and then sent to a CAEN DT5740 digitizer with a sampling rate of 62.5 MS/s and 12-bit resolution~\cite{Alekseev:2016llm}. The signal charge was reconstructed either by integrating the waveform (and normalizing to the number of photoelectrons) or by using the peak amplitude. The signal timing was extracted using the time at which the signal reached 50\% of its peak amplitude (constant fraction method).

\begin{figure*}
\centering
\begin{tabular}{cc}
\includegraphics[width=0.35\textwidth]{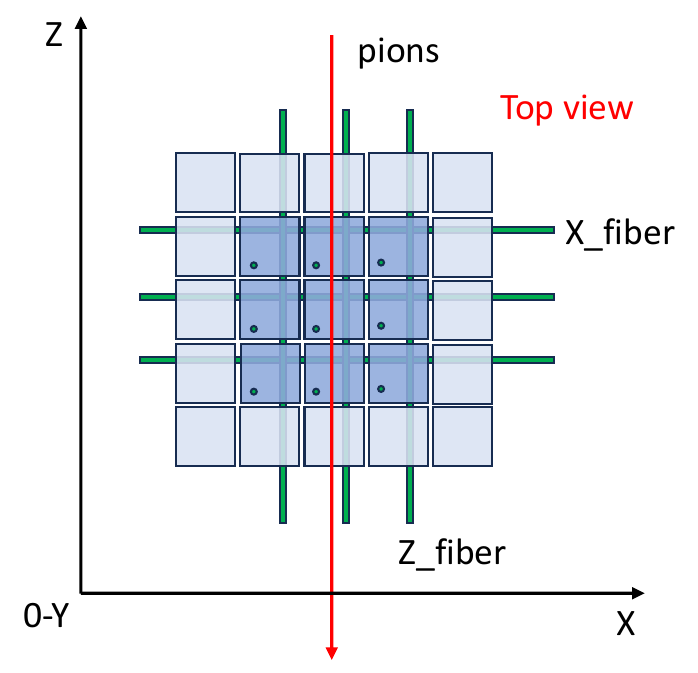}
&
\includegraphics[width=0.65\textwidth]{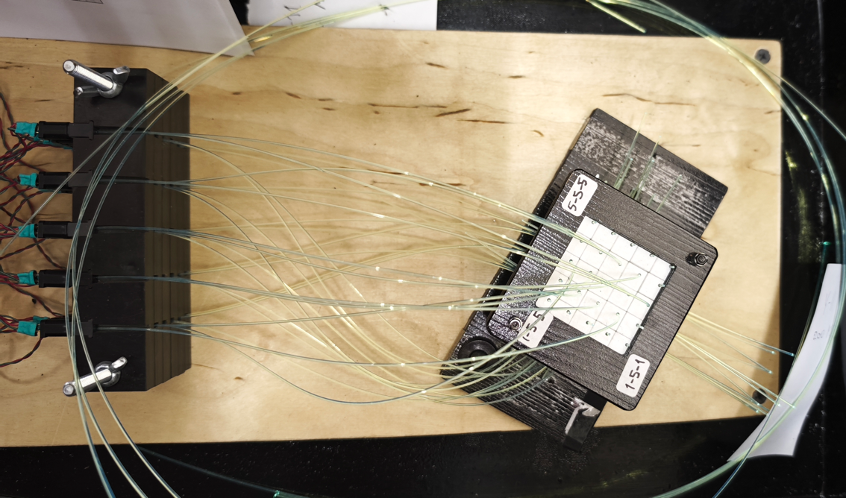}
\end{tabular}
\caption{Left: readout of 27 WLS fibers by the digitizer; inactive fibers are not shown. Right: top view of the full test setup.}
\label{setup_inside_box}
\end{figure*}

In total, 27 WLS fibers were read out by the digitizer, as shown in Figure~\ref{setup_inside_box}. Each fiber went through a row of 5 cubes. For these cubes, signals from three orthogonal fiber directions (X, Y, and Z) were available, allowing us to study the light collection from three sides of each selected cube. For clarity, we assigned a unique label (from 0 to 26) to each cube in the central array. Channel numbers were defined according to the electronics connection map.


The beam test was carried out at the SC-1000 proton synchrocyclotron at the B.P. Konstantinov Petersburg Nuclear Physics Institute in Gatchina, Russia. The 730 MeV/$c$ $\pi^{-}$ beam was limited in intensity to $10^{4}$ particles per second. An iron collimator suppressed the beam halo and constrained the beam size to 150 mm in the horizontal projection.

\begin{figure*}
\centering
\includegraphics[width=0.85\textwidth]{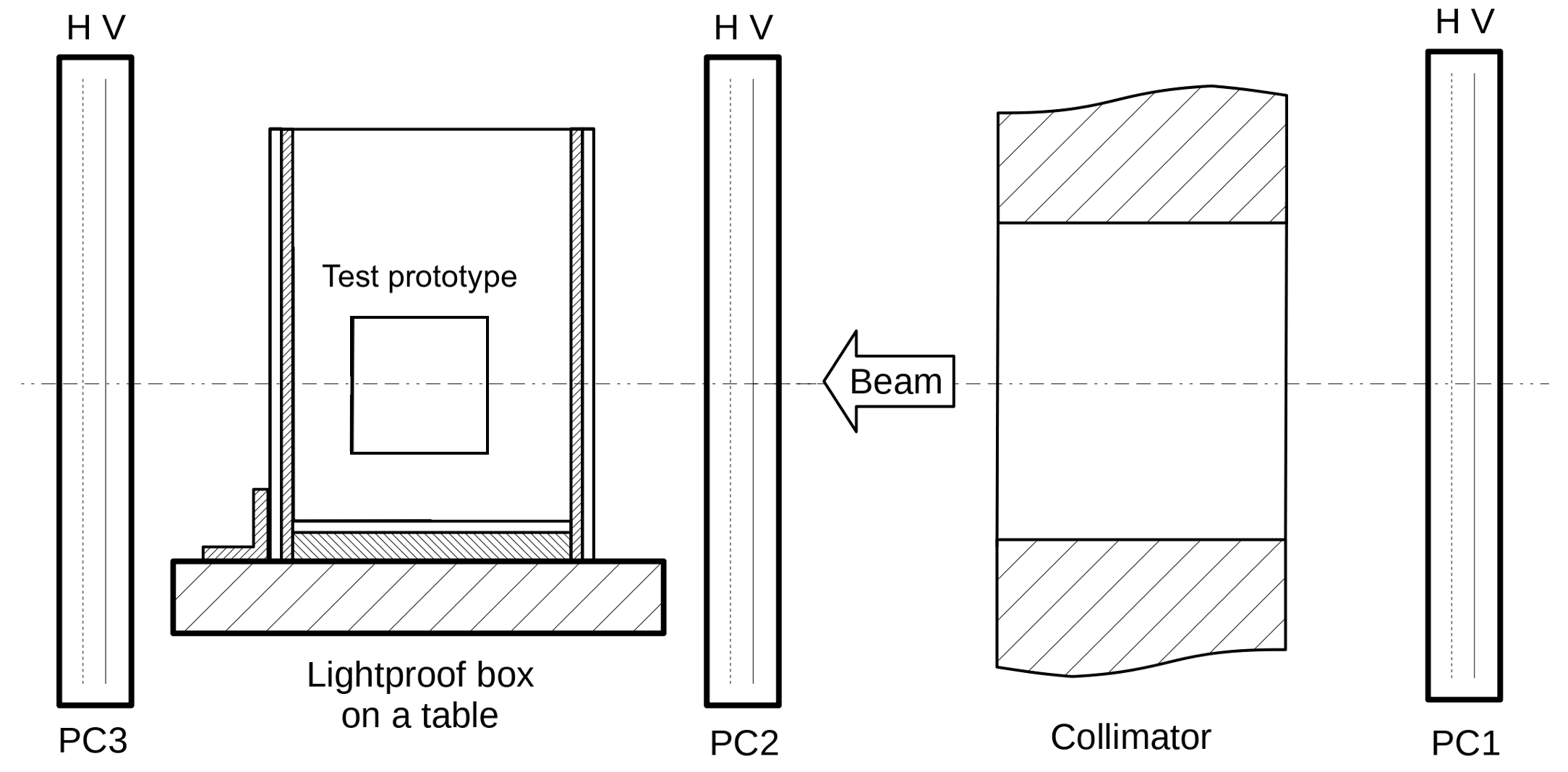}
\caption{Test setup at the pion beam line 1 of the PNPI synchrocyclotron.}
\label{fig:lanl-setup}
\end{figure*}

 The SFGD prototype was positioned on a table between a system of three proportional chambers (PC1–PC3) (see Figure~\ref{fig:lanl-setup}). Each chamber was equipped with sensitive wires in both horizontal (H) and vertical (V) directions, with a wire pitch of 1 mm. The tracking efficiency of each chamber exceeded 99\%. The particle trajectory was reconstructed using hits in all three chambers. Track reconstruction accuracy (of about 0.5 mm) in the region of the studied scintillator  cubes  is determined mainly by the multiple scattering in the trigger counter and cubes themselves. Only events with straight-line tracks were selected for analysis.

\label{sec:setup}
\section{Signal waveform analysis}
In this section, we describe the procedure for reconstructing key signal parameters from the digitized SiPM waveforms recorded in the SuperFGD prototype. The waveforms contain detailed information about the light signals produced in scintillator cubes, and allow for the precise  extraction of observables such as light yield and timing. To obtain these quantities, each recorded waveform was analyzed to determine three key parameters: signal amplitude, integrated charge (proportional to the number of photoelectrons), and time of arrival.

\begin{figure*}
\centering
\begin{tabular}{cc}
\includegraphics[width=0.45\textwidth]{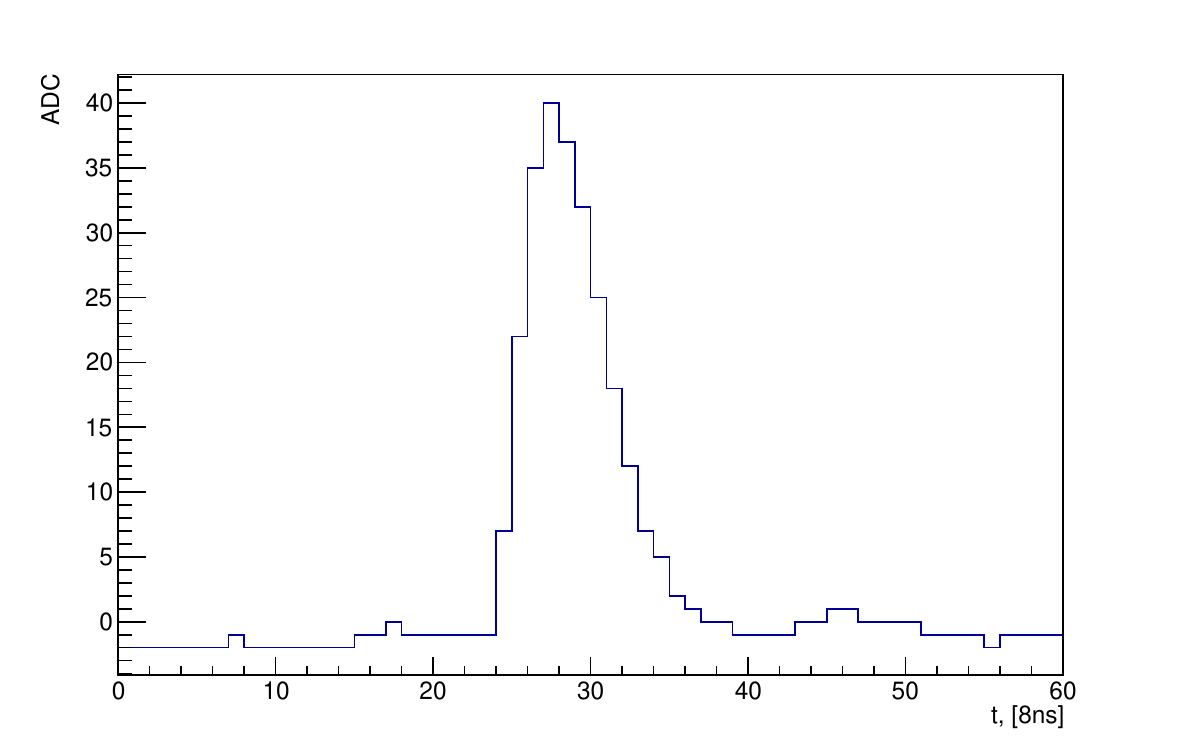}
&
\includegraphics[width=0.45\textwidth]{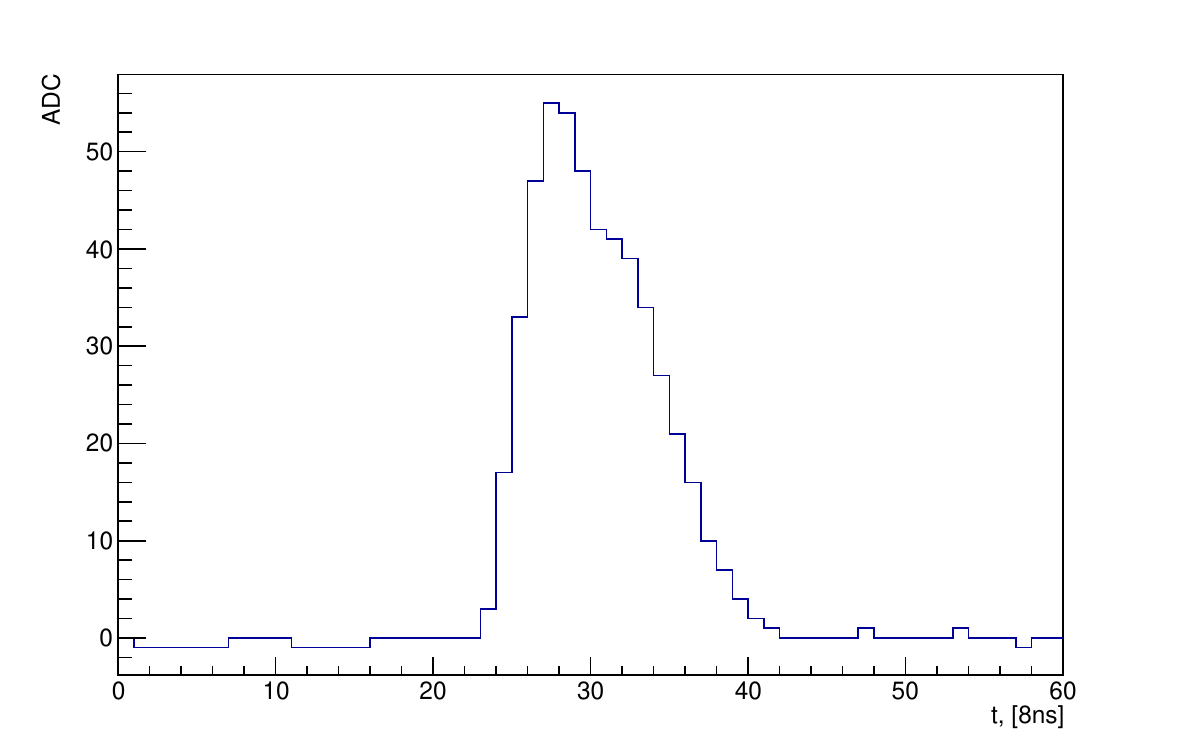}
\end{tabular}
\caption{Examples of digitized SiPM signals. Left: clean single pulse. Right: waveform with afterpulsing or pile up with accidentals.}
\label{fig:signal_1}
\end{figure*}

Figure~\ref{fig:signal_1} left shows an example of typical SiPM waveforms. For each waveform, the following quantities were computed:
\begin{itemize}
\item \textbf{Amplitude}: peak value of waveform after baseline subtraction;
\item \textbf{Baseline}: the average level in the pre-pulse region;
\item \textbf{Integrated charge}: the integral of the waveform in a fixed time window of $25 \times 8 = 200$~ns, starting from the pulse threshold.
\end{itemize}

\begin{figure*}
\centering
\begin{tabular}{cc}
\includegraphics[width=0.45\textwidth]{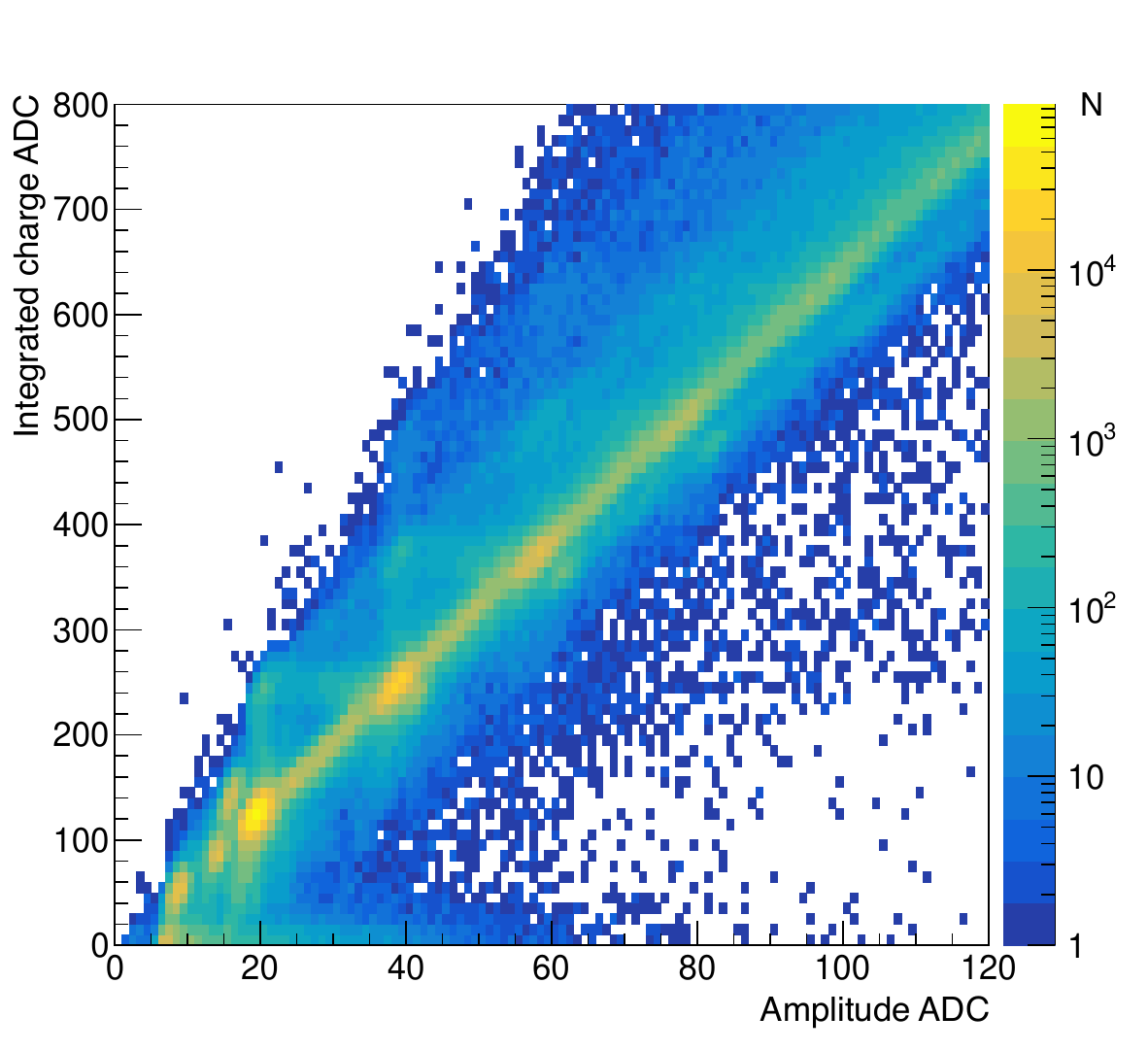}
&
\includegraphics[width=0.45\textwidth]{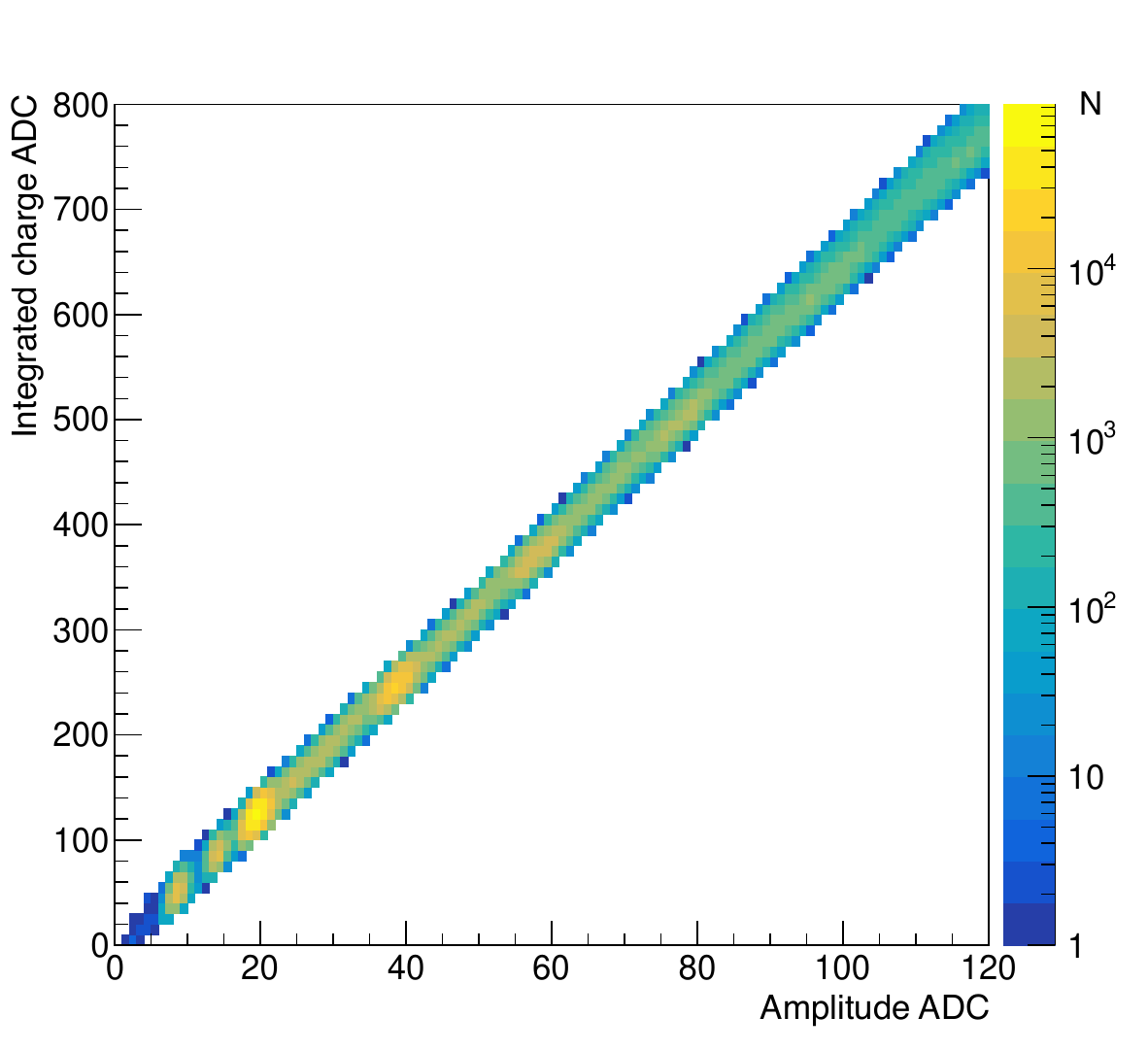}
\end{tabular}
\caption{Left: Amplitude vs. integrated charge for calibration events. The diagonal band corresponds to clean single pulses. Right: Off-diagonal events are rejected.}
\label{fig:calibration_signals_cut}
\end{figure*}

One in every 5000 self-triggered events was saved for analysis and SiPM calibration. Events with multiple pulses in the integration window were rejected. This was done using a two-dimensional cut on the amplitude vs. charge distribution: single-pulse events form a diagonal band (see Figure~\ref{fig:calibration_signals_cut}). Events outside this band, especially those with low amplitude and high charge, were attributed to overlapping pulses or noise and excluded. This selection was primarily applied to self-triggered calibration events.

\begin{figure*}
\centering
\begin{tabular}{cc}
\includegraphics[width=0.45\textwidth]{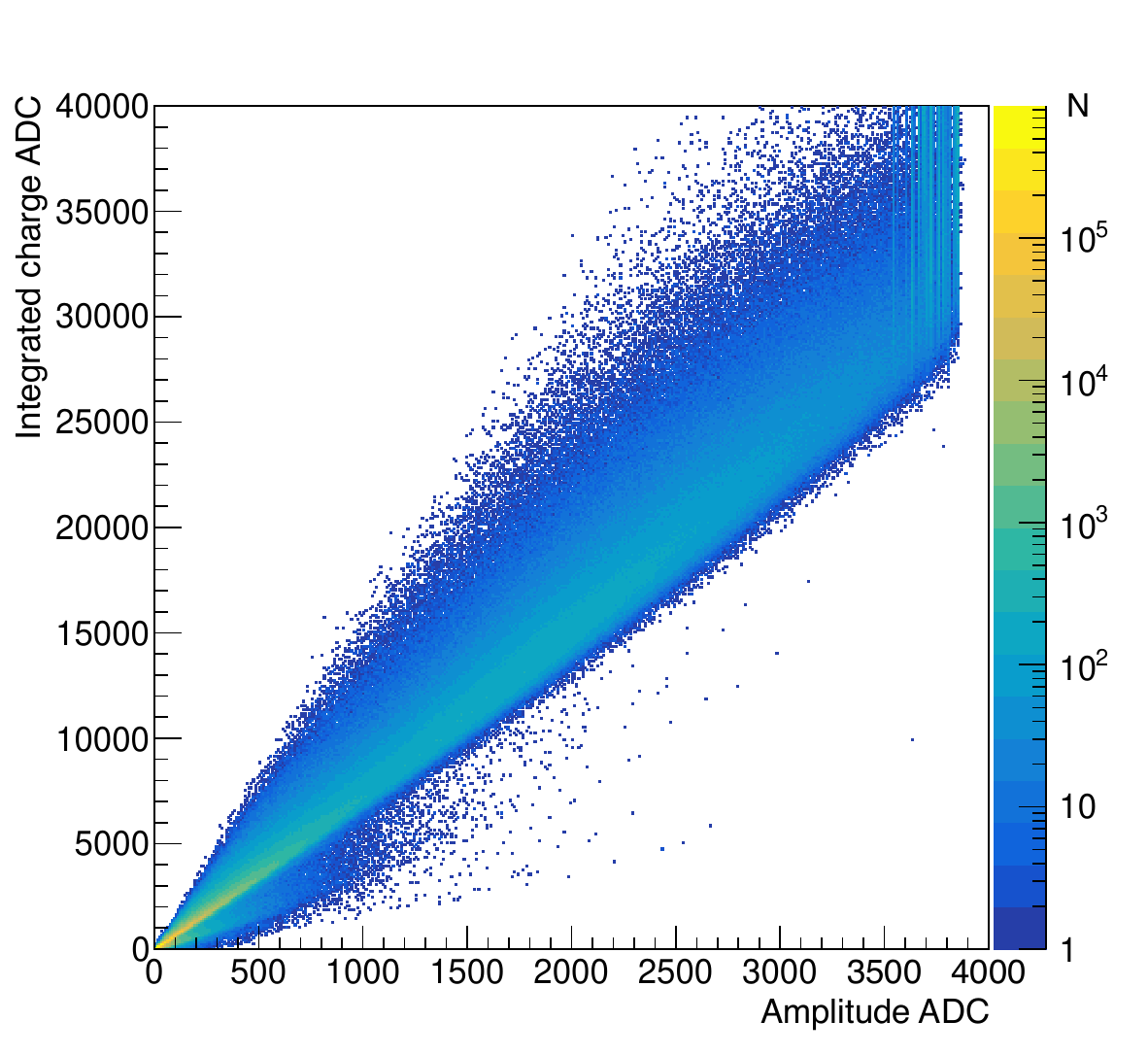}
&
\includegraphics[width=0.45\textwidth]{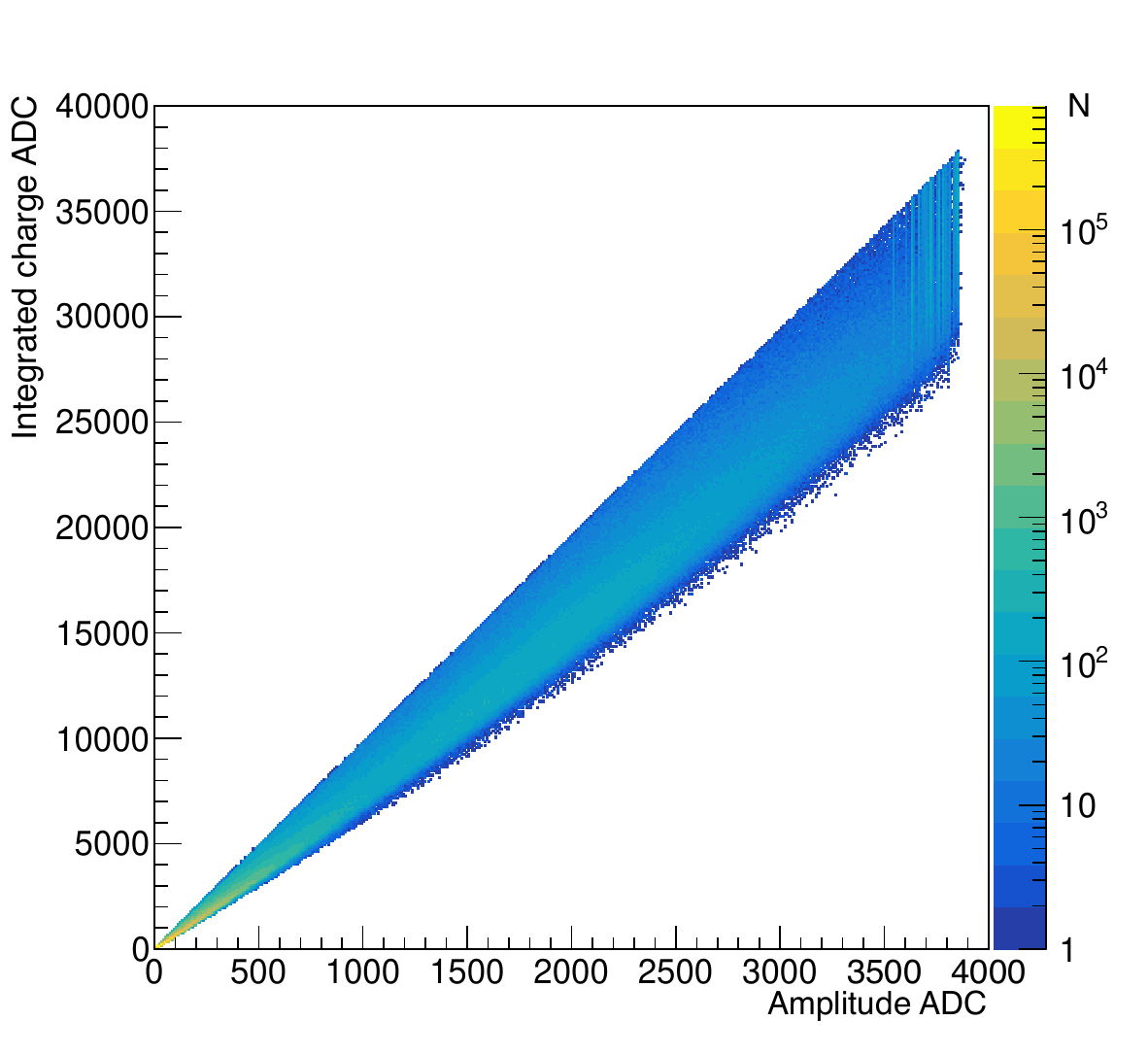}
\end{tabular}
\caption{Left: Amplitude vs. integrated charge for beam-triggered events. Right: A looser selection is applied.}
\label{fig:events_signals_cut}
\end{figure*}

For beam-triggered events (pions), more complex waveforms were observed. A relaxed selection cut was used to retain valid signals even if they were affected by moderate pile-ups (see Figure~\ref{fig:events_signals_cut}).

Some waveforms suffered from signal saturation due to the limited ADC range (typically in the range of 3500 -- 4000 ADC counts depending on the channel). To reconstruct the true amplitude and charge for these overflowed signals, we implemented a dedicated waveform fitting procedure using an analytical function. The fitting was performed on the non-saturated portion of the waveform using the following empirical model:

\begin{equation}
\label{eq:fit_signal}
\begin{split}
\xi &= \frac{x - t_0}{\tau \cdot n}\,, 
\\
F(x) &= A \cdot \exp \Big( n \cdot \big(1 + \log(\xi) - \xi \big) \Big) + B\,,
\end{split}
\end{equation}

\noindent where:

\begin{itemize}
\item $A$ -- amplitude,
\item $t_0$ -- pulse start time,
\item $B$ -- baseline,
\item $n$ and $\tau$ -- shape parameters.
\end{itemize}

\begin{figure*}
\centering
\begin{tabular}{cc}
\includegraphics[width=0.45\textwidth]{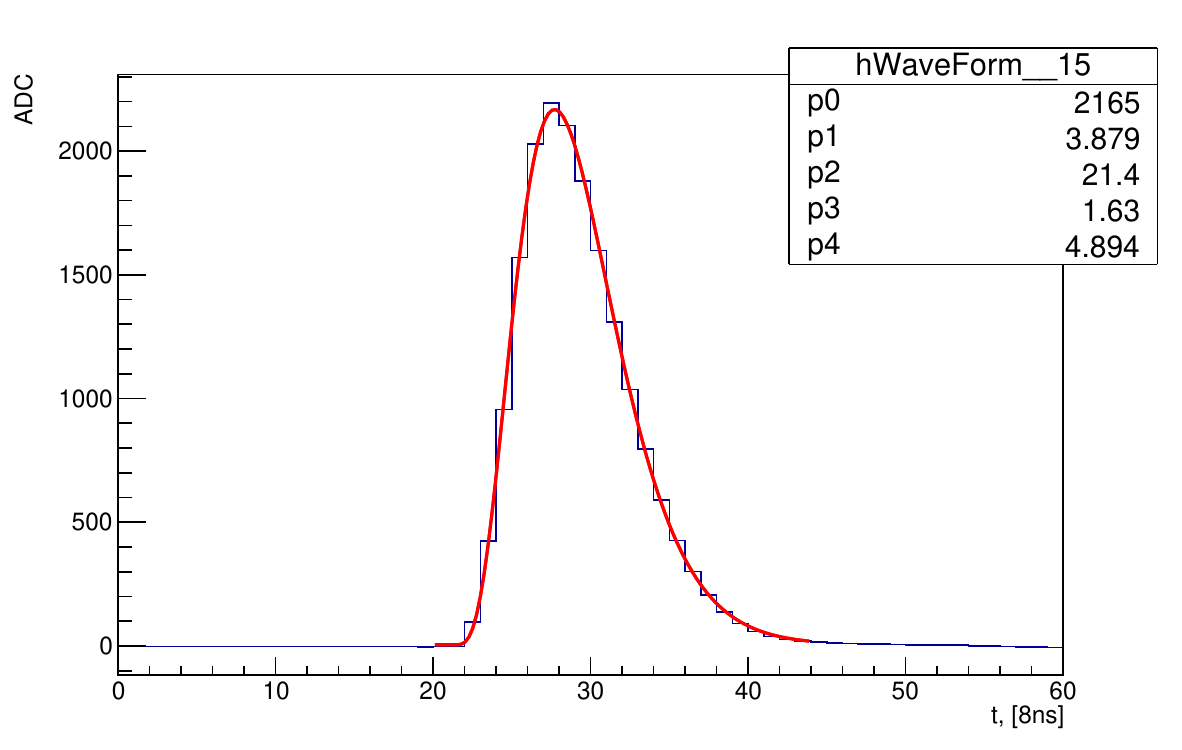}
&
\includegraphics[width=0.45\textwidth]{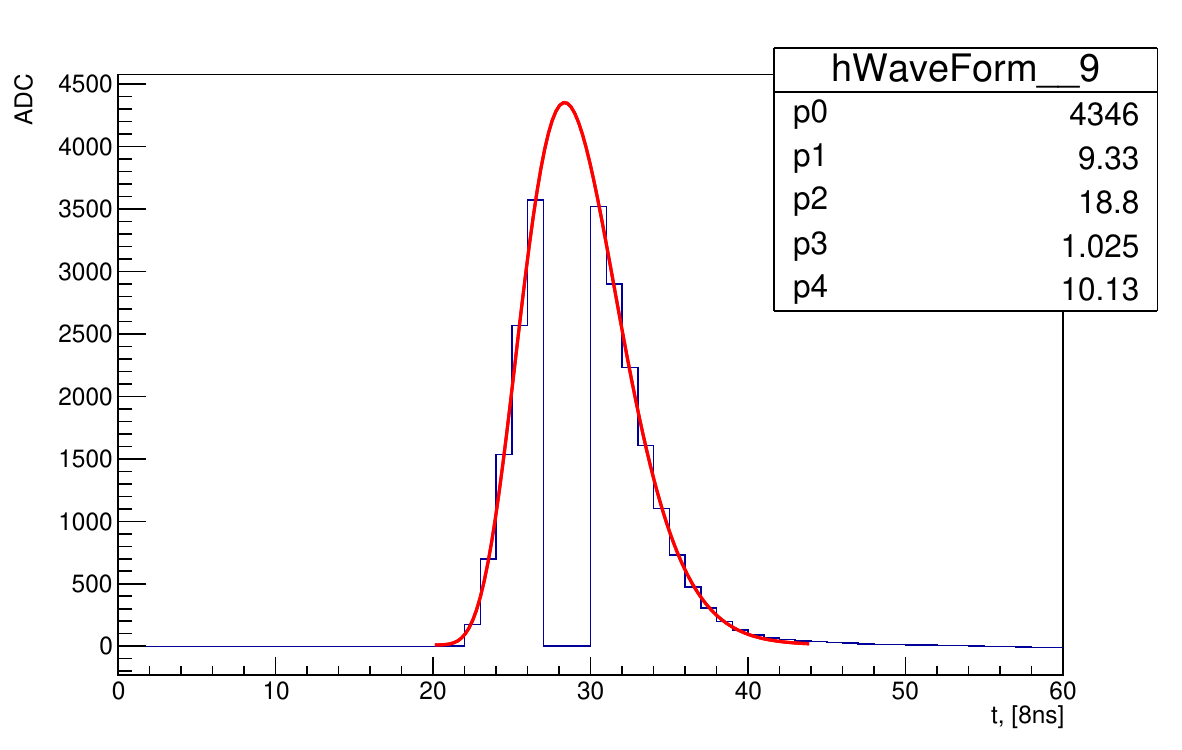}
\end{tabular}
\caption{Waveform fit examples. Left: standard signal. Right: saturated signal with excluded overflow region.}
\label{fig:signals_fit}
\end{figure*}

\begin{figure*}
\centering
\includegraphics[width=0.7\textwidth]{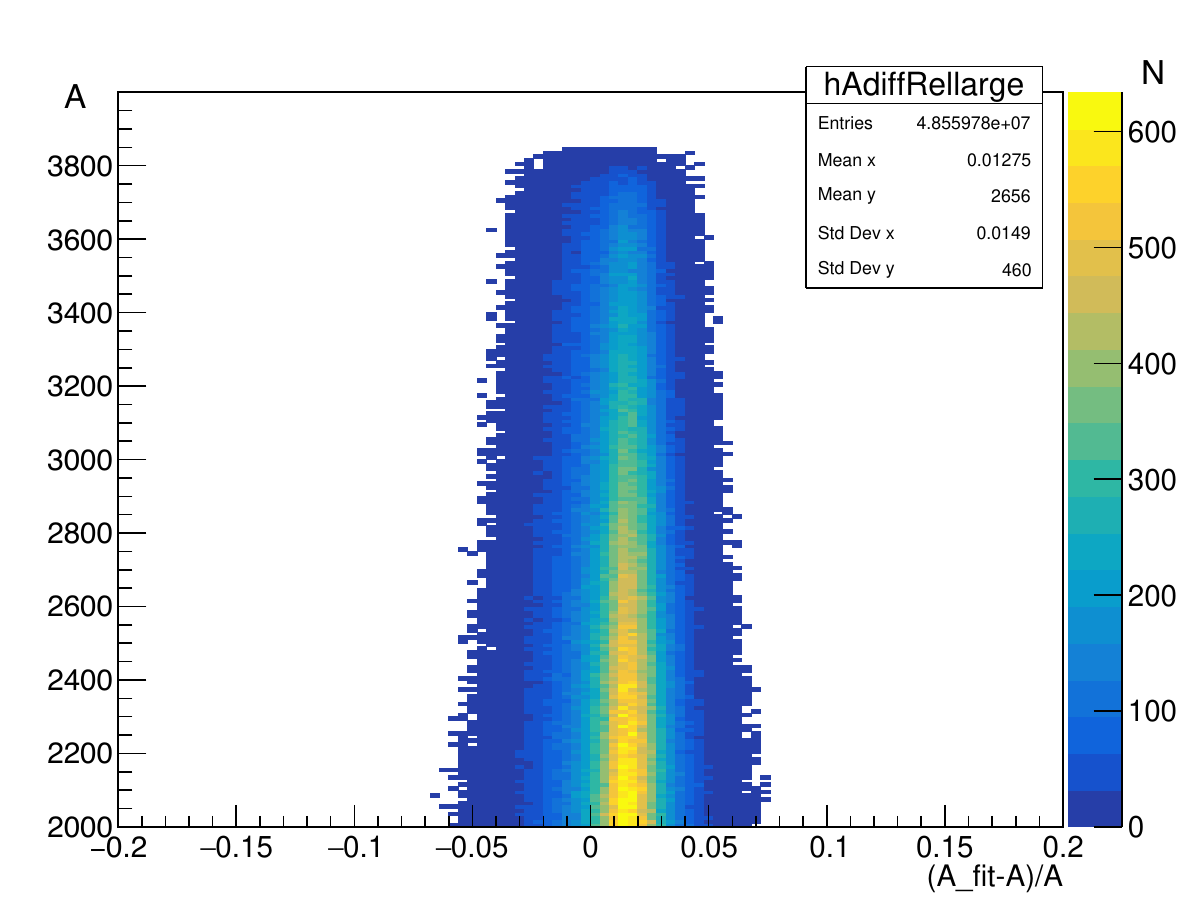}
\caption{Relative difference between fitted and measured peak amplitudes: $(A_{\text{fit}} - A)/A$ for non-saturated signals.}
\label{fig:fit_performance}
\end{figure*}

Examples of waveform fits are shown in Figure~\ref{fig:signals_fit}. Saturated parts of the waveform were excluded from the fit. To assess the fit performance, the fitted amplitude $A_{\text{fit}}$ was compared with the actual measured amplitude $A$ for non-saturated pulses. The relative difference $(A_{\text{fit}} - A)/A$ is shown in Figure~\ref{fig:fit_performance}.

The reconstructed signal parameters -- peak amplitude, charge, and arrival time -- were used as inputs for further analyses of the SuperFGD prototype performance, including light yield and optical crosstalk.

\label{sec:data_flow}
\section{SiPM calibration}
To accurately interpret the signals produced in the scintillator cubes, a calibration of the photosensors is required. Silicon photomultipliers exhibit gain and crosstalk characteristics that must be quantified to convert signal amplitude and charge into physical quantities such as the number of photoelectrons (p.e.).

In this section, we describe the procedure used to extract the SiPM gain and optical crosstalk probability from waveform data. The SiPM calibration data were collected during the same runs as the beam data, but in the intervals between beam spills. By studying the distribution of signal amplitudes and integrated charges for events with different number of photoelectrons, we determined the conversion factors and validated the linearity of the response. These results are essential for subsequent measurements of light yield and signal uniformity.

During these intervals, the digitizer was triggered on SiPM dark counts, and signals were selected as described in Section~\ref{sec:data_flow}. Peak amplitude data were used in the calibration procedure. For each readout channel, the following parameters were extracted: the SiPM gain, the pedestal position, and the optical crosstalk probability.

\begin{figure*}
\centering
\begin{tabular}{cc}
\includegraphics[width=0.47\textwidth]{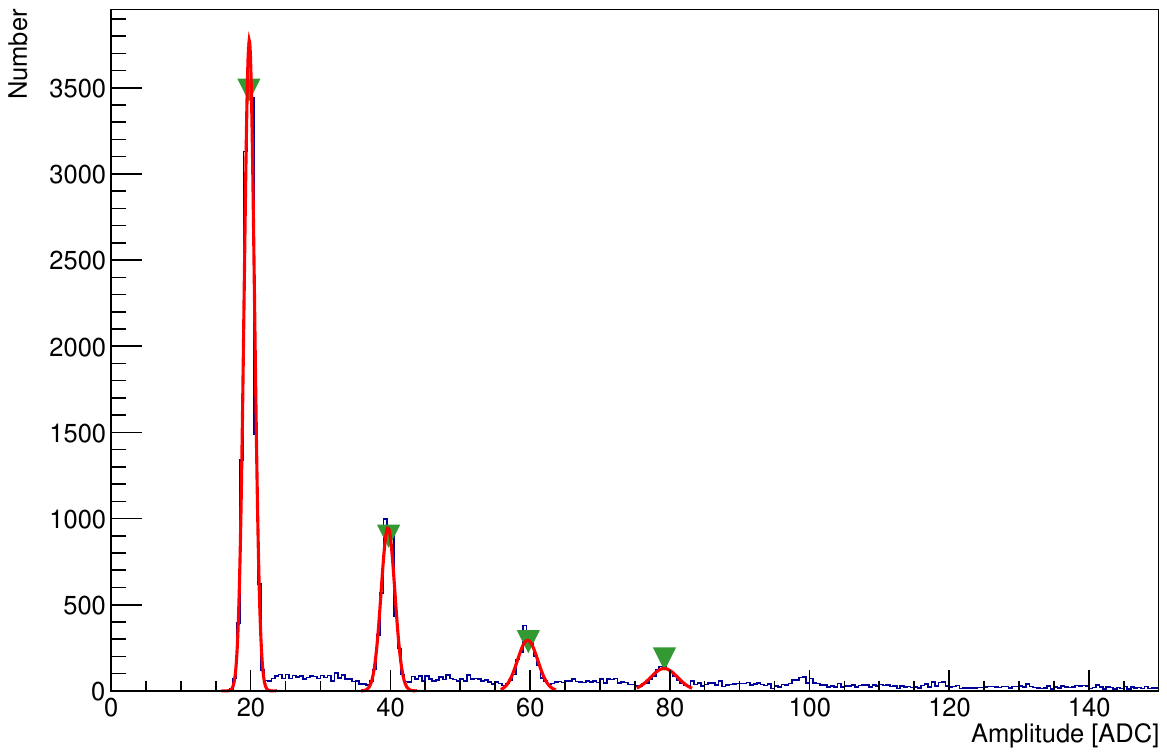}
&
\includegraphics[width=0.47\textwidth]{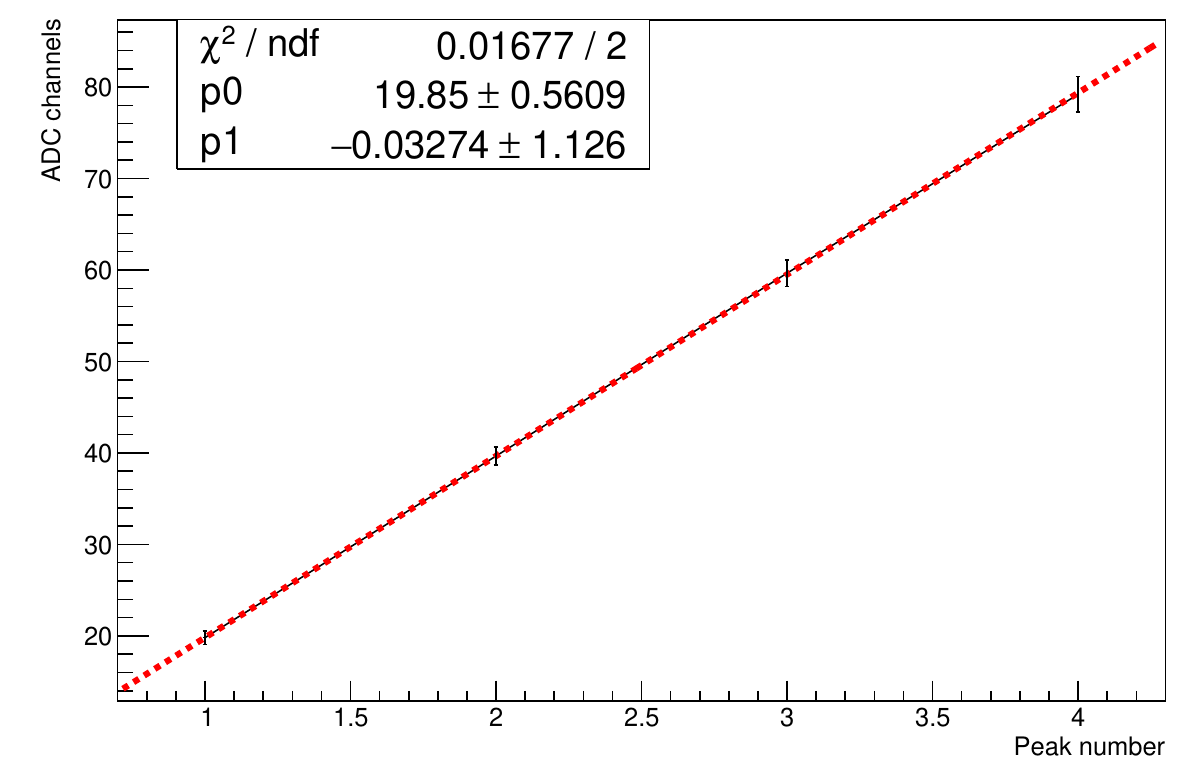}
\end{tabular}
\caption{Peak amplitude distribution from SiPM dark noise signals: (left) raw data including background; (right) the distances between adjacent peaks were then fitted with a linear function, and the slope was interpreted as the SiPM gain. Green triangles mark the peak positions identified using the \texttt{TSpectrum} algorithm.}
\label{fig:calibration_Amplitude}
\end{figure*}

The gain extraction was measured using the CERN ROOT framework~\cite{rene_brun_2019_3895860}. Individual photoelectron peaks were identified using the \texttt{TSpectrum} peak-searching algorithm. Each peak was fitted with a Gaussian function to determine its position and width. Examples of the amplitude distribution is shown in Figure~\ref{fig:calibration_Amplitude}~(left). The distances between adjacent peaks were then fitted with a linear function, and the slope was interpreted as the SiPM gain as shown in Figure~\ref{fig:calibration_Amplitude}~(right).

The optical crosstalk, defined as the probability of simultaneous firing of multiple microcells in response to a single photoelectron, was estimated for each channel as:
\begin{equation}
\text{crosstalk} = \frac{\overline{A} - A_{\text{pedestal}}}{A_{1\text{pe}}} - 1,  
\end{equation}
where $\overline{A}$ is the average signal amplitude, $A_{\text{pedestal}}$ is the pedestal level, and $A_{1\text{pe}}$ is the mean position of the first p.e. peak. This estimate assumes that the mean signal amplitude is increased due to contributions from correlated avalanches (optical photons emitted during a Geiger discharge triggering neighboring cells). A similar procedure was applied using the integrated charge distribution to provide a cross-check.

The extracted calibration constants (gain, pedestal, and optical SiPM crosstalk) were stored for each channel and used in the reconstruction and analysis of physics events.
\label{sec:Calibration}
\section{Calculating the position of scintillator cubes}
Before performing the fiber or cube LY analysis, it was necessary to determine the physical position of each individual cube.

For each cube, an event map was constructed by selecting events in which all three associated readout channels (X, Y, and Z) registered signals above a threshold of 5 photoelectrons to suppress electronic noise and reduce contributions from low-energy background. This procedure yielded 27 individual event maps (one per cube) each built independently based on the corresponding readout geometry. All histograms were created with identical binning (bin size is 0.5~mm) and coordinate systems, based on the pion track positions reconstructed with the tracking chambers.

\begin{figure*}
\centering
\begin{tabular}{ccc}
\includegraphics[width=0.33\textwidth]{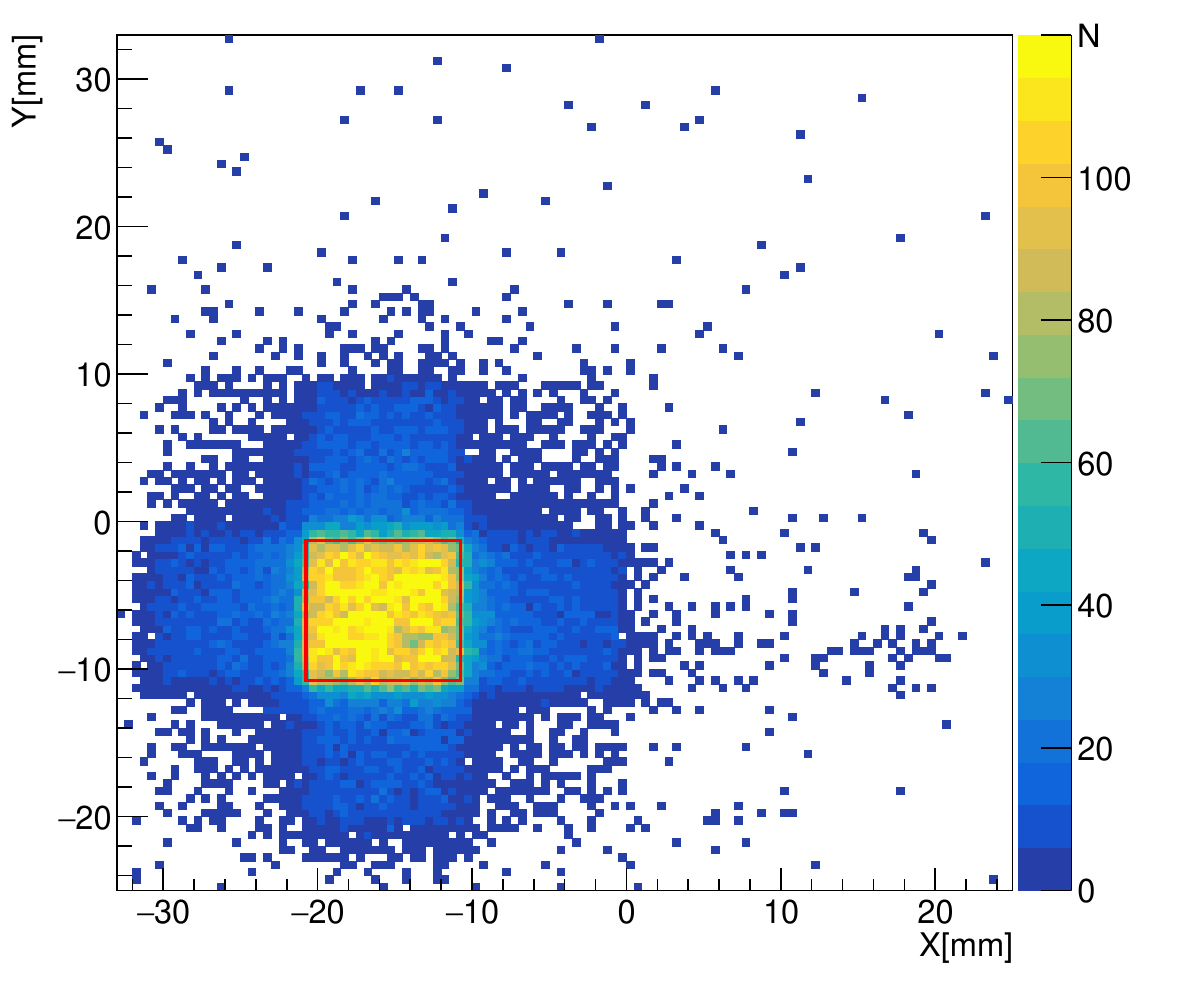}
&
\includegraphics[width=0.33\textwidth]{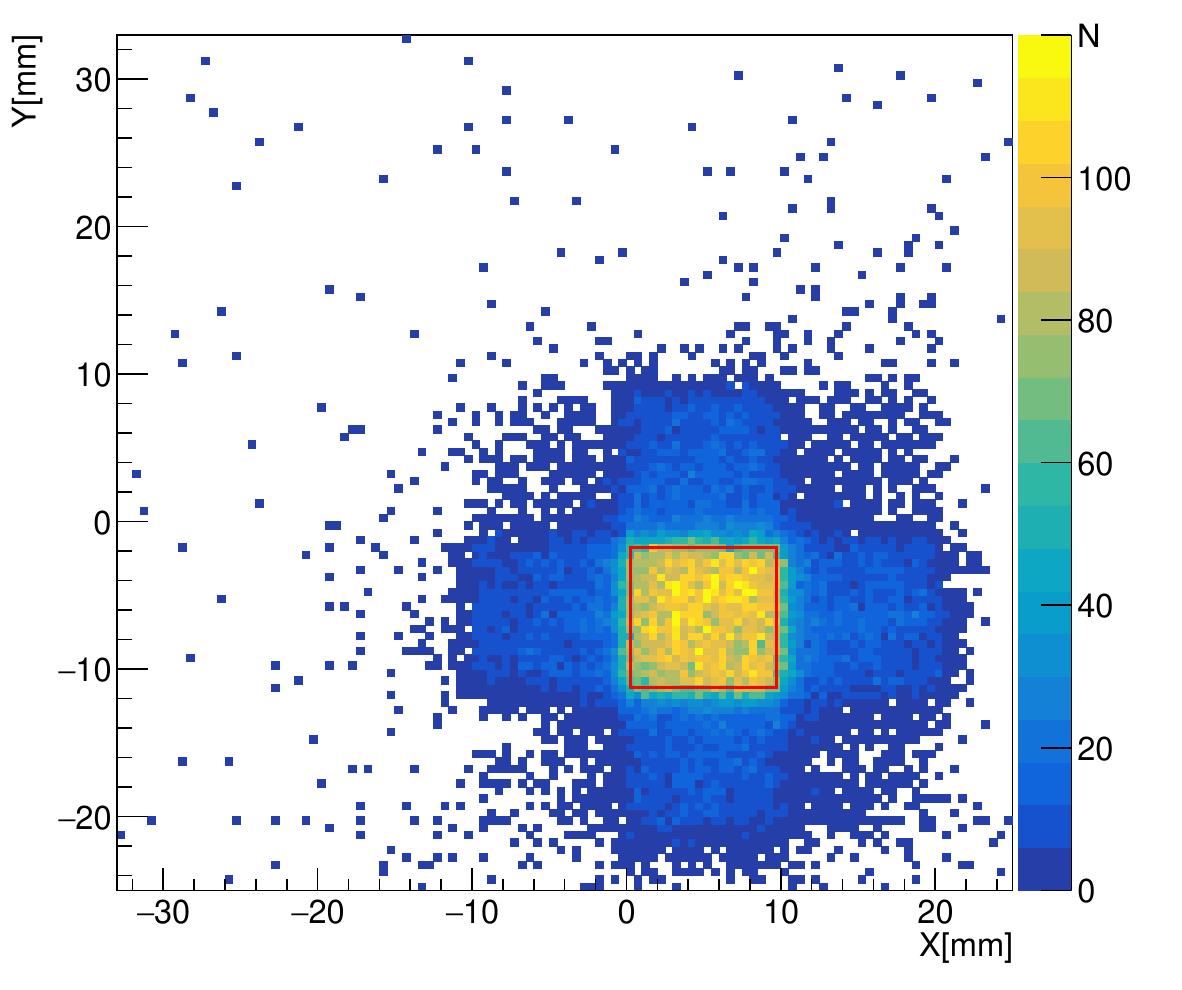}
&
\includegraphics[width=0.33\textwidth]{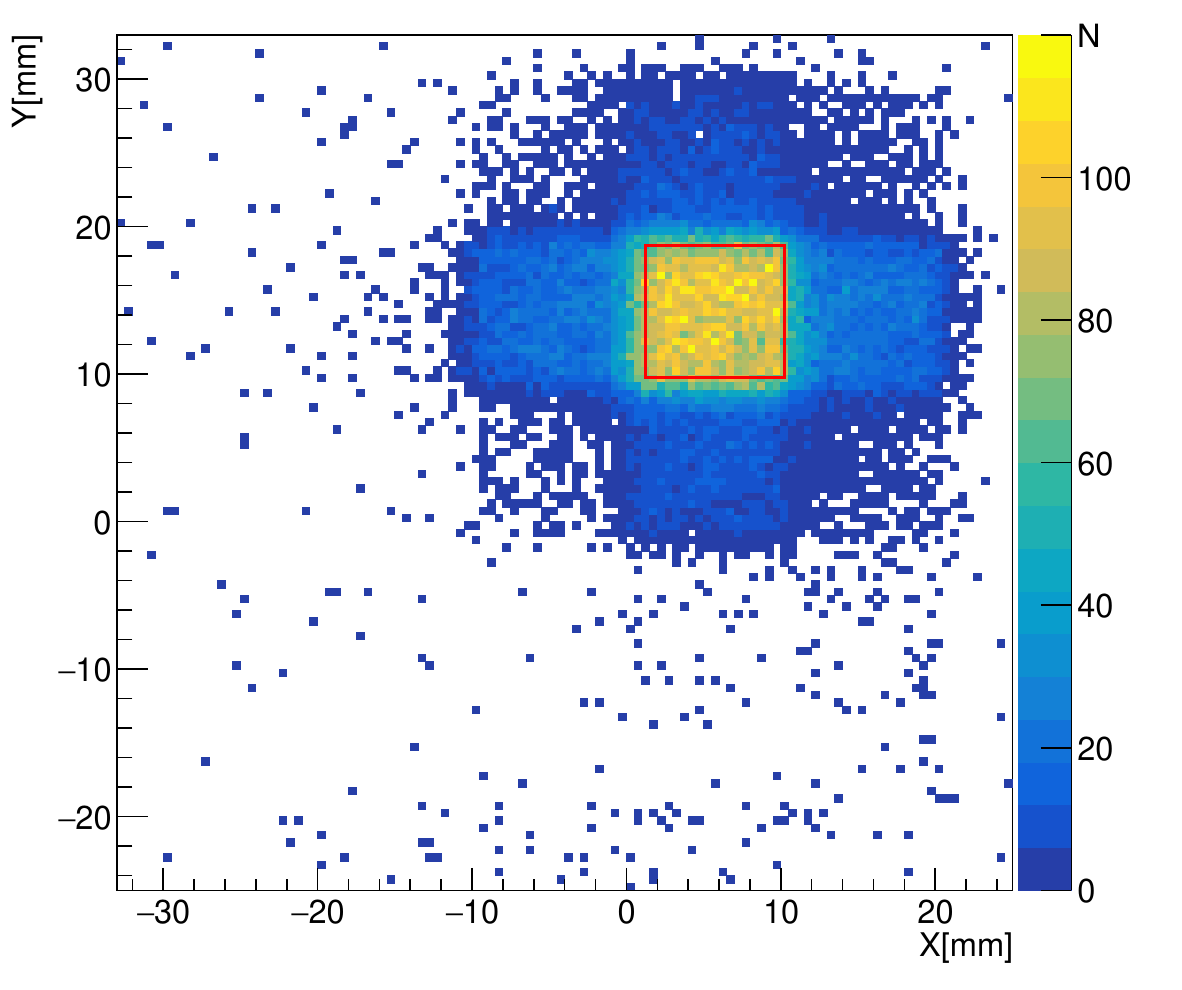}
\end{tabular}
\caption{Reconstructed hit position maps for cubes 0, 5, and 26 in the $(X, Y)$ plane. The maps show event densities based on the pion track reconstruction. The red box indicates the reconstructed cube boundaries, defined as the smallest axis-aligned rectangle enclosing all bins with occupancy above a threshold of 80 entries per bin.}
\label{fig:cubes_position}
\end{figure*}

Figure~\ref{fig:cubes_position} shows examples of reconstructed event distributions in the $(X, Y)$ plane for cubes 0, 5, and 26. The coordinates represent the interpolated positions of pion tracks at the center of the plane. Only events corresponding to straight-line tracks and simultaneous signals from all three associated fibers were included.

The color scale indicates the number of events per bin. Clear clusters are observed, corresponding to the geometrical positions of the cubes within the detector volume. The boundaries of each cube were determined using a data-driven approach based on the event density distribution. In particular, all bins with occupancy above a threshold of 80 entries per bin were identified, and the cube boundary was defined as the smallest axis-aligned rectangle enclosing all such bins. The resulting boundary is therefore always rectangular by construction, and is shown as a red box in Figure~\ref{fig:cubes_position}.

In addition to the central peak, a diffuse area with lower occupancy (typically below 40 events per bin) can be observed around some cubes. These regions are attributed to the optical crosstalk between neighboring cubes and will be discussed in more detail in Section~\ref{sec:xTalk1D}.

This approach provides robustness against statistical fluctuations and detector resolution effects while maintaining a well-defined geometrical region for further analysis. This procedure allows us to suppress low-occupancy regions while preserving a consistent geometrical definition of the cube size. The resulting boundaries are shown as red contours in Figure~\ref{fig:cubes_position}.

This position identification procedure was applied uniformly to all 27 cubes under analysis. The use of a common binning scheme and alignment guarantees consistent treatment across the dataset, enabling quantitative studies of detector uniformity and signal response.

\label{sec:Cubes_position}
\section{Light Yield from each fiber}
A key parameter for evaluating the performance of the detector is the LY, defined as the number of p.e. detected per traversing minimum ionizing particle (MIP). Accurate determination of the light yield is essential for assessing the efficiency and uniformity of the readout system, as well as for validating the overall scintillation and light collection performance.

The analysis was performed using the beam data and applying the SiPM gain calibration described in Section~\ref{sec:Calibration}. For each cube, we selected only those events where the reconstructed track passed through the cube volume, as defined in Section~\ref{sec:Cubes_position}. For each selected hit, the signal amplitude, integrated charge, and reconstructed charge from a waveform fit were independently converted into the number of photoelectrons using the formula:

\begin{equation}
\label{eq:LY_no_overflow}
LY_{\text{p.e.}} = \frac{LY_{\text{ADC}} - \text{pedestal}}{\text{gain} \times (1 + \text{crosstalk})},
\end{equation}

In cases where the waveform was saturated (overflowed) and a fit was used for amplitude recovery, a small bias correction factor $R_{\text{fit}}$ was introduced to match the fitted amplitude to the actual scale:

\begin{equation}
\label{eq:LY_with_fit}
LY_{\text{p.e.}} = \frac{LY_{\text{ADC}} - \text{pedestal}}{\text{gain} \times (1 + \text{crosstalk}) \times R_{\text{fit}}},
\end{equation}

\noindent where $R_{\text{fit}} = 1.013$ was obtained from the residual difference between real and fitted amplitudes (Section~\ref{sec:data_flow}, the mean x value from Figure~\ref{fig:fit_performance}).



\begin{figure*}
\centering
\begin{tabular}{ccc}
\includegraphics[width=0.33\textwidth]{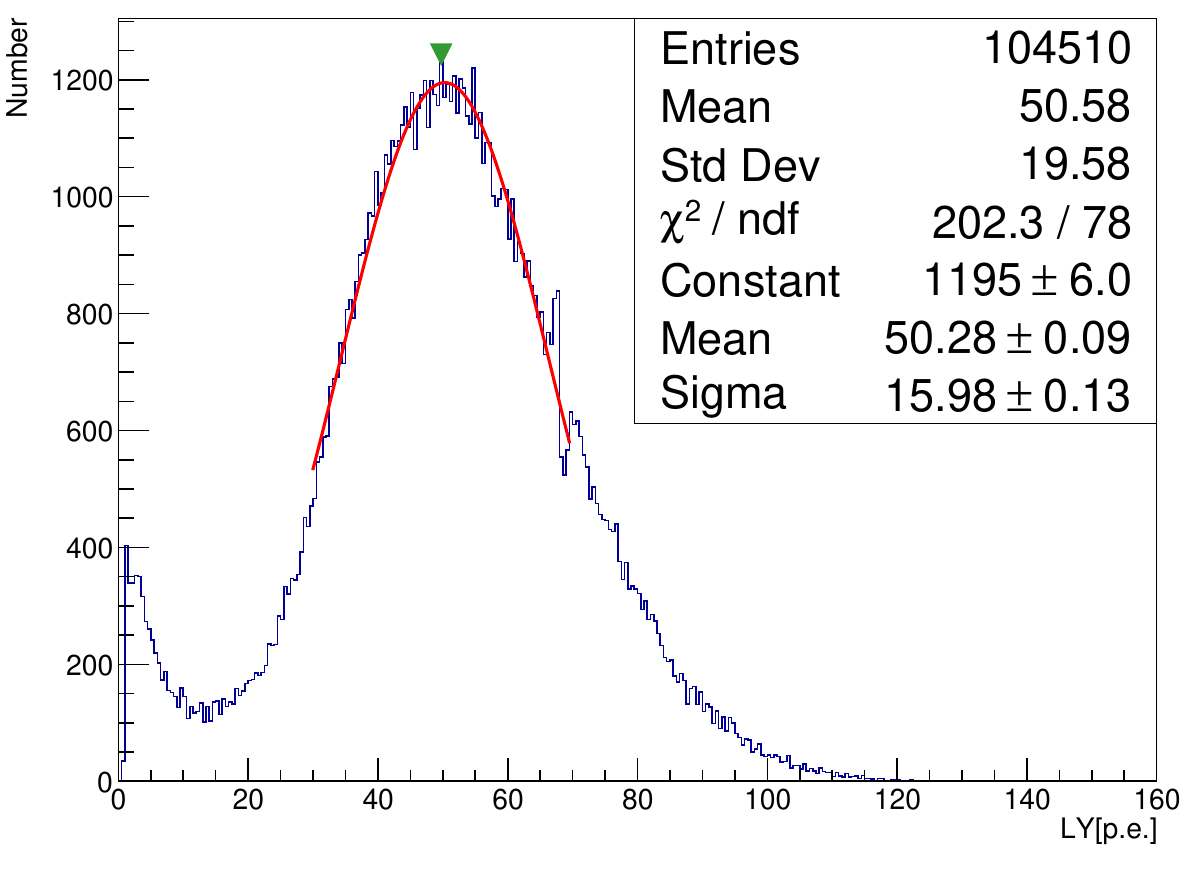}
&
\includegraphics[width=0.33\textwidth]{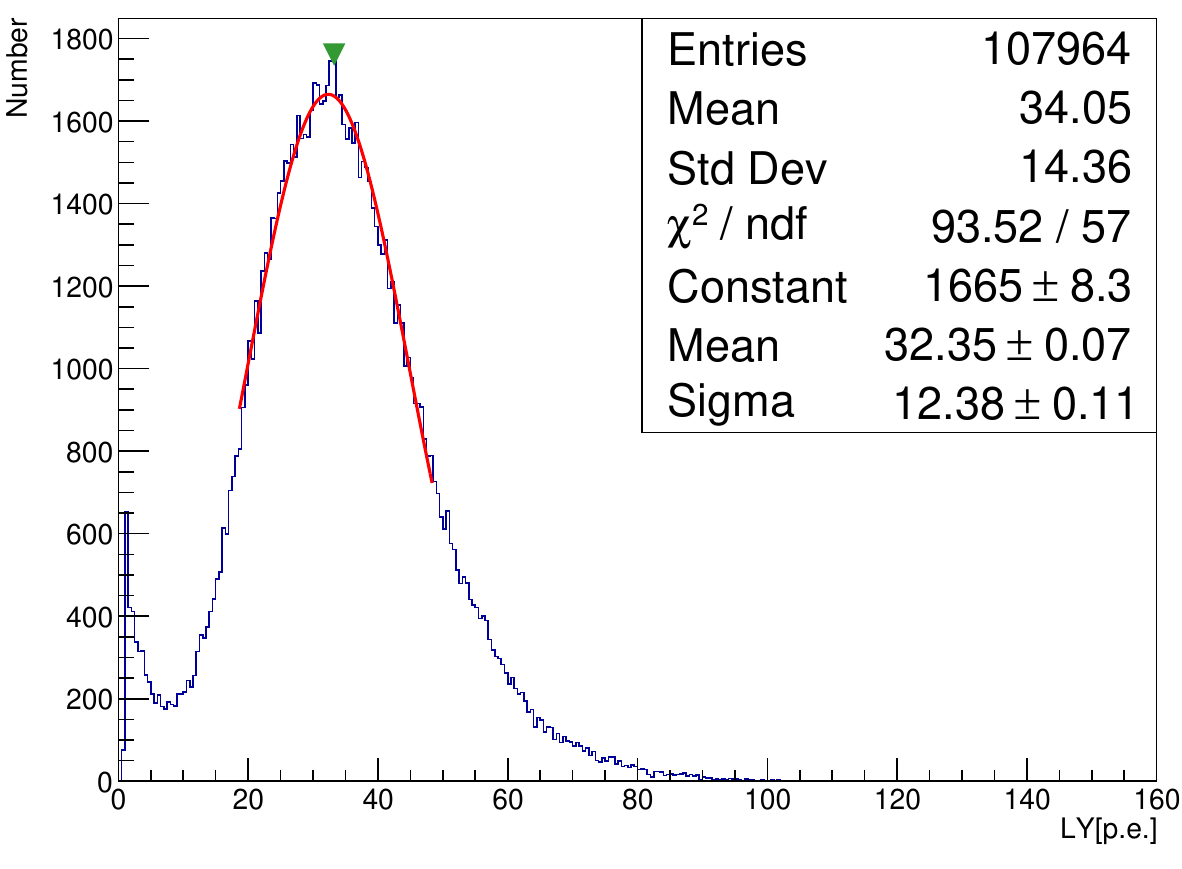}
&
\includegraphics[width=0.33\textwidth]{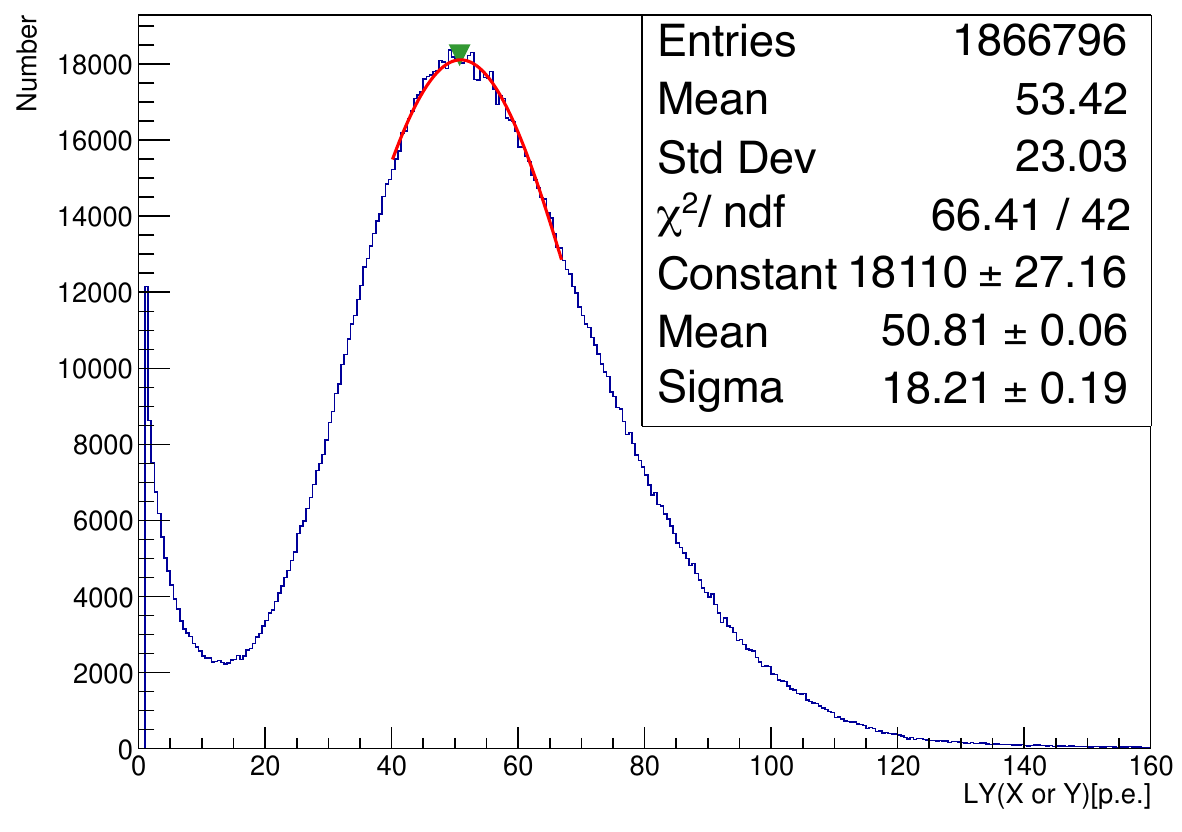}
\end{tabular}
\caption{Light yield distributions for cubes connected to channel 6 (left), channel 21 (center), and all channels combined (right). The variation is primarily attributed to fiber-SiPM coupling differences. Green triangle marks the peak position identified using the \texttt{TSpectrum} algorithm.}
\label{fig:different_LY}
\end{figure*}

During the analysis, we observed notable variations in light yield between different channels. These differences can be attributed to variations in fiber-to-SiPM coupling quality and mechanical alignment. Figure~\ref{fig:different_LY} illustrates this effect by showing the average light yield distributions for cubes connected to channel 6 (left), channel 21 (center), and all channels combined (right).

To reduce the impact of variations in fiber-to-SiPM coupling, a normalization procedure was applied. For each readout channel, the light yield distribution was fitted with a Gaussian function, and the mean of the fit was used as a normalization factor. The response of each channel was then scaled relative to the global average light yield, which was itself obtained from a Gaussian fit to the combined distribution of all channels. All light yield results presented in the following sections include this normalization, enabling consistent comparison across channels and ensuring uniformity in detector response analysis.

This normalization procedure enables channel-to-channel comparison and consistent detector-level characterization. It also allows for identification of systematic effects and the evaluation of signal response uniformity across the instrumented volume.
\label{sec:Fiber_LY}
\section{Light Yield Distribution Inside Scintillator Cubes}
To evaluate the uniformity of light collection within individual scintillator cubes, we studied the spatial dependence of the light yield as a function of the particle track position. Using high-resolution tracking information from the beam chambers, we reconstructed the pion position in the cubes central plane with about 0.5~mm precision. The selected events were binned within each cube using a spatial granularity of 0.5~mm in the $X$ and $Y$ directions.

For each spatial bin, we calculated the average light yield using signals from different combinations of readout fibers:
\begin{itemize}
\item The sum of the signals from the $X$ and $Y$ fibers (with $X$ denoting the horizontal fiber, and $Y$ the vertical fiber),
\item The signal from the $X$ fiber alone,
\item The signal from the $Y$ fiber alone.
\end{itemize}

This analysis enables us to characterize the uniformity of the light response, detect potential asymmetries in light collection, and investigate how the signal varies with the particle track’s position relative to the fiber orientation. Results are presented for several representative cubes and compared between readout directions.

To extract the position-dependent light yield, we employed the same binning (0.5~mm) and coordinate system as described in Section~\ref{sec:Cubes_position}. For each bin, the light yield distribution was fitted with a Gaussian function, and the mean of the fit was used as the representative value. This fitting-based approach reduces sensitivity to statistical fluctuations and improves robustness against noise.

\begin{figure*}
\centering
\begin{tabular}{cc}
\includegraphics[width=0.455\textwidth]{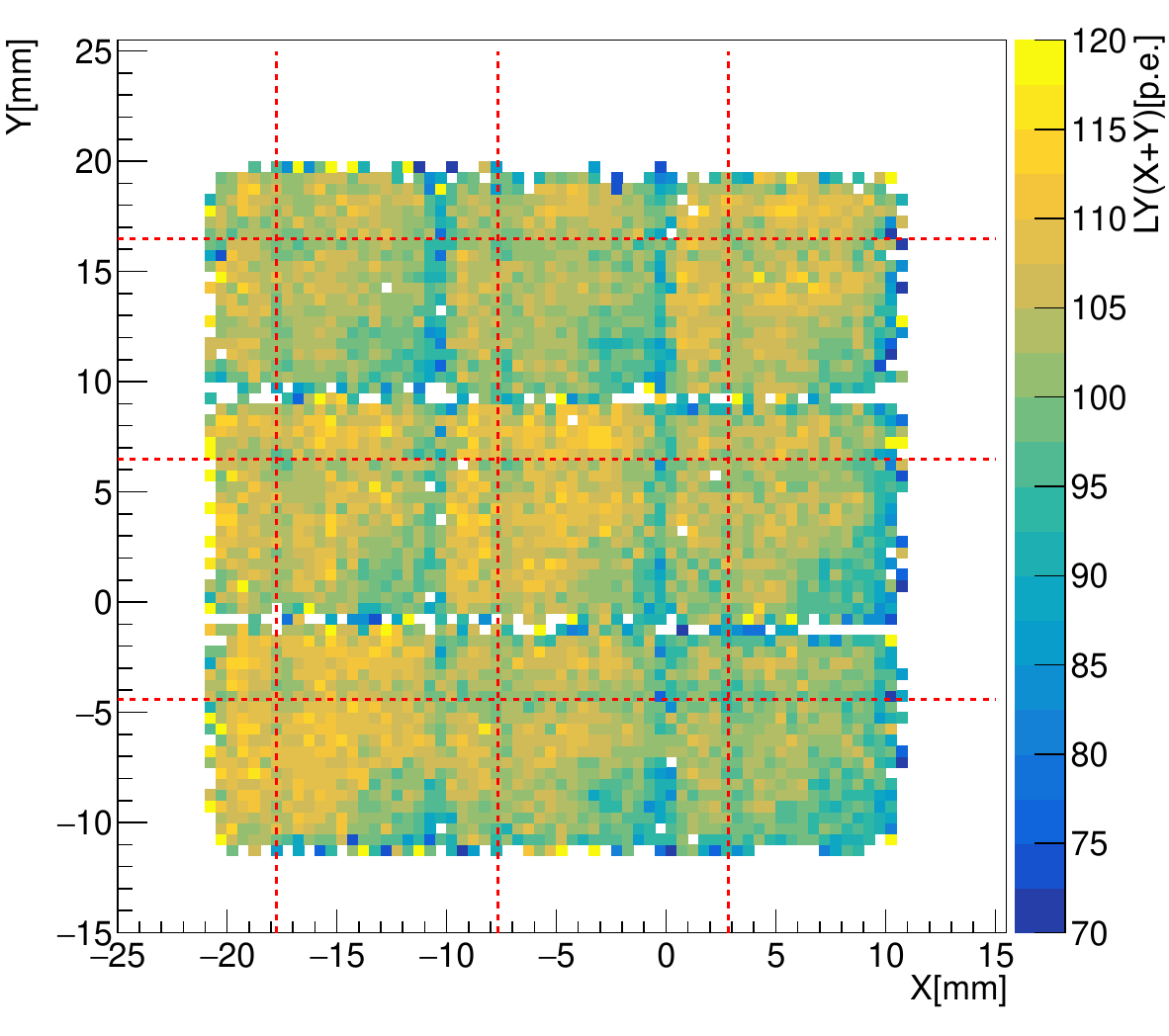}
&
\includegraphics[width=0.41\textwidth]{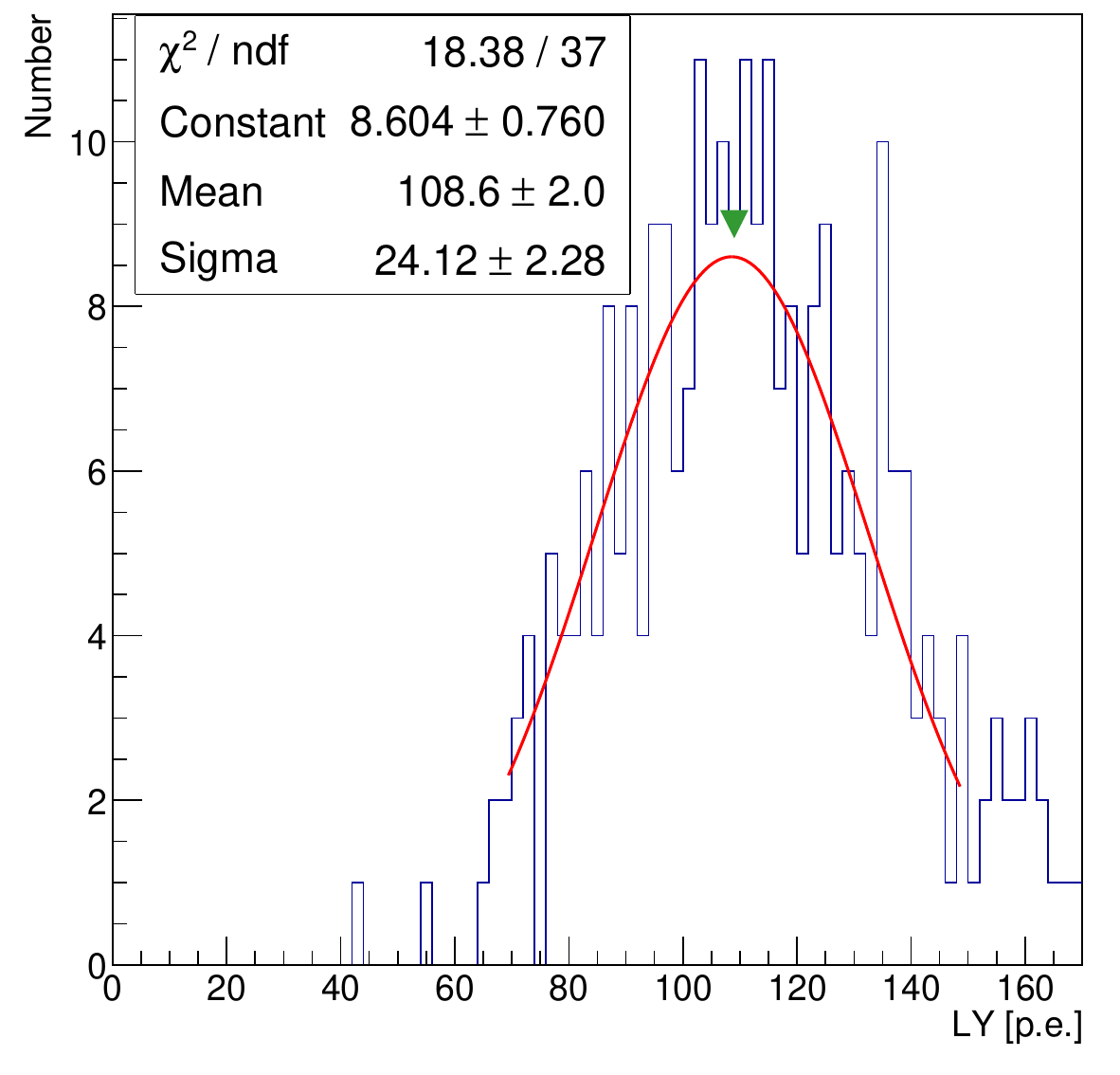}
\end{tabular}
\caption{Left: position-dependent light yield obtained from the sum of $X$ and $Y$ fibers signals for a $3\times3$ cube region. The red dotted lines indicate the positions of the $X$-oriented and $Y$-oriented fibers. Right: example of a light yield distribution in a single spatial bin, fitted with a Gaussian function. Green triangle marks the peak position identified using the \texttt{TSpectrum} algorithm.}
\label{fig:LY_XplusY}
\end{figure*}

Figure~\ref{fig:LY_XplusY} shows a two-dimensional light yield map for a $3\times3$ group of cubes, using the sum of the signals from both $X$ and $Y$ fibers. The response is approximately symmetric, with a peak in the central region and gradual attenuation toward the edges of cubes. This map represents the most complete light collection scenario, and it serves as a benchmark for evaluating the performance of individual fibers.

\begin{figure*}
\centering
\begin{tabular}{cc}
\includegraphics[width=0.455\textwidth]{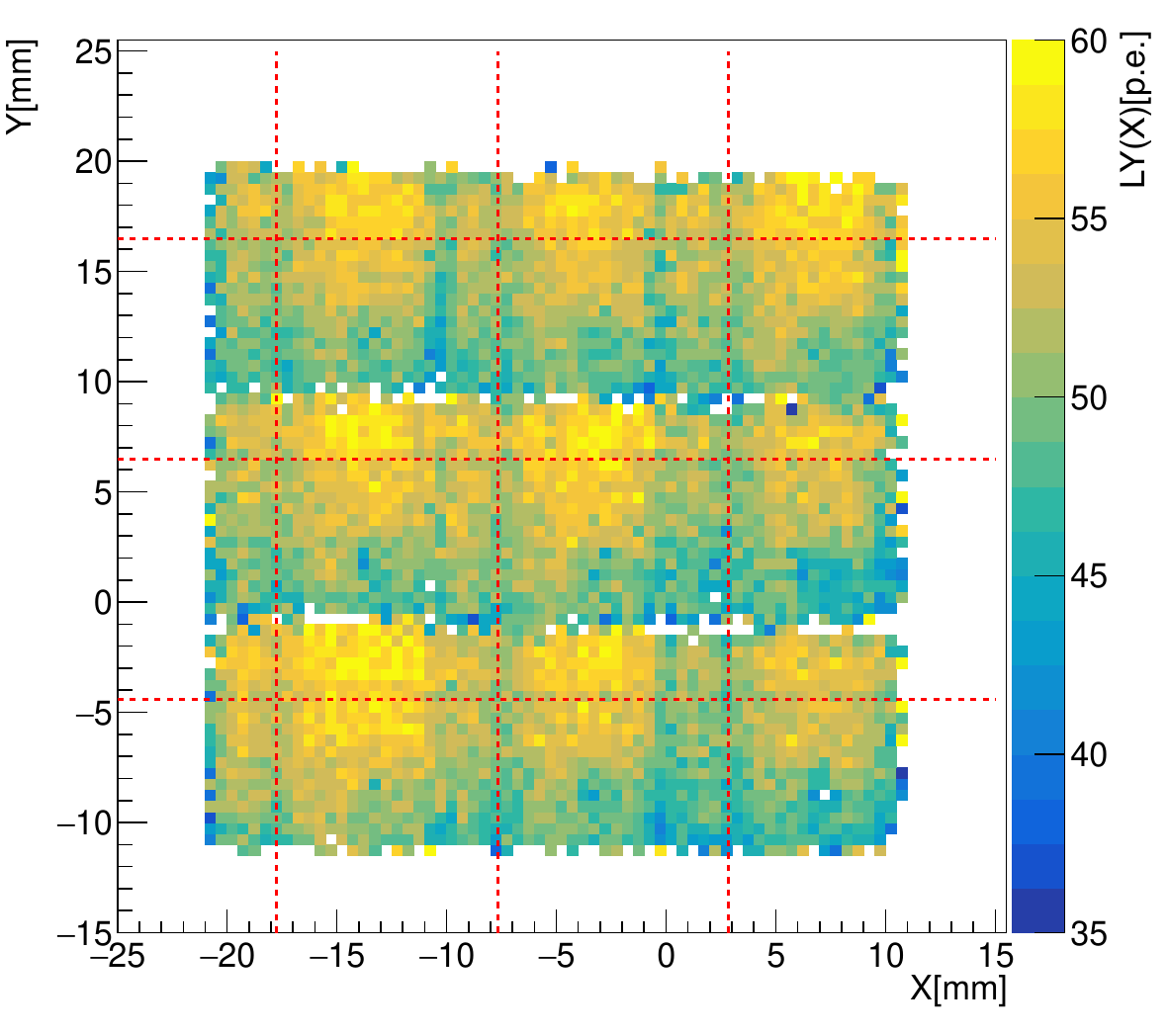}
&
\includegraphics[width=0.41\textwidth]{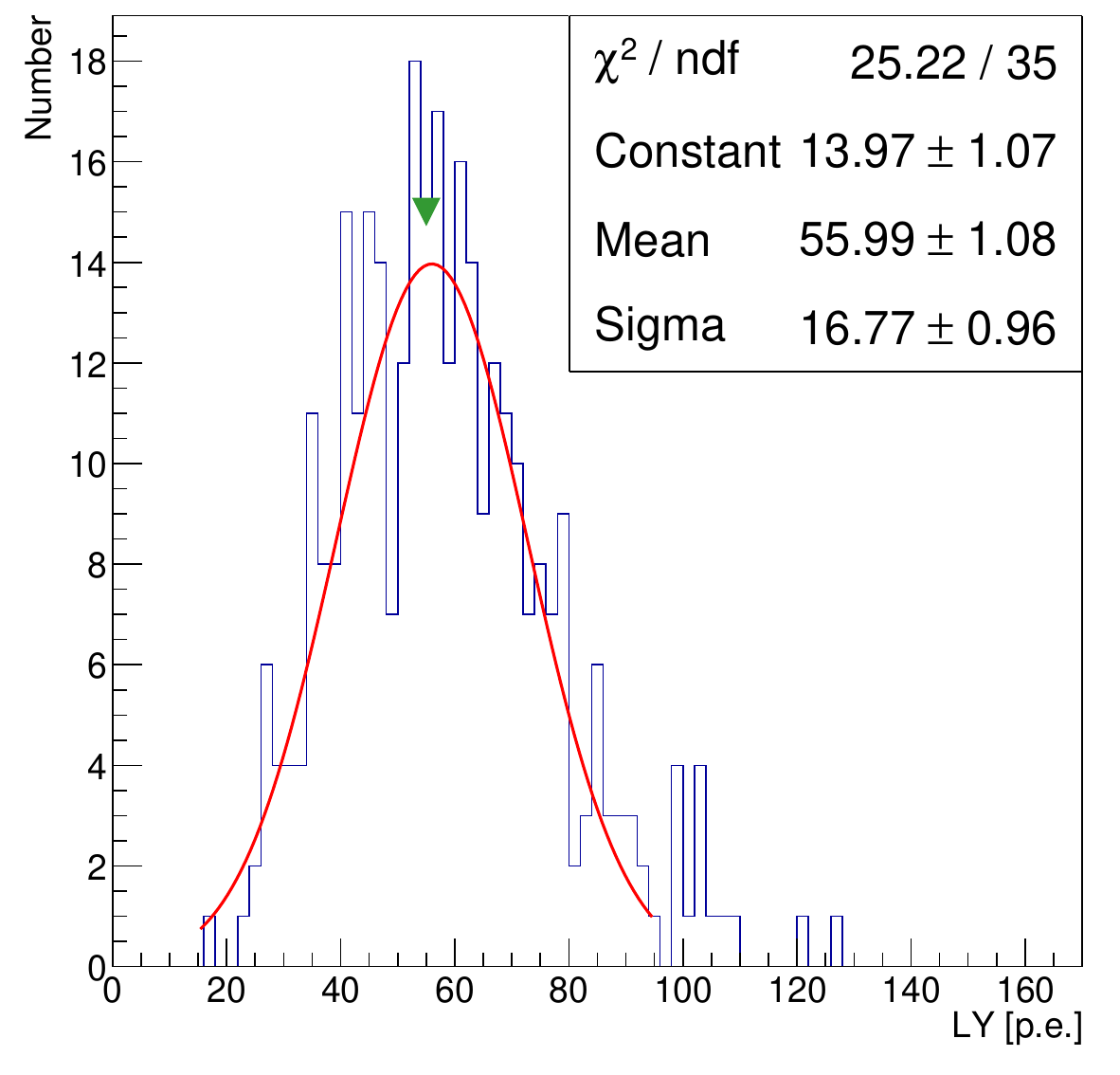}
\end{tabular}
\caption{Left: position-dependent light yield from $X$-oriented fibers only. The red dotted lines indicate the positions of the $X$-oriented and $Y$-oriented fibers. Right: example of a Gaussian fit to the light yield distribution in a single bin. Green triangle marks the peak position identified using the \texttt{TSpectrum} algorithm.}
\label{fig:LY_Xonly}
\end{figure*}

\begin{figure*}
\centering
\begin{tabular}{cc}
\includegraphics[width=0.455\textwidth]{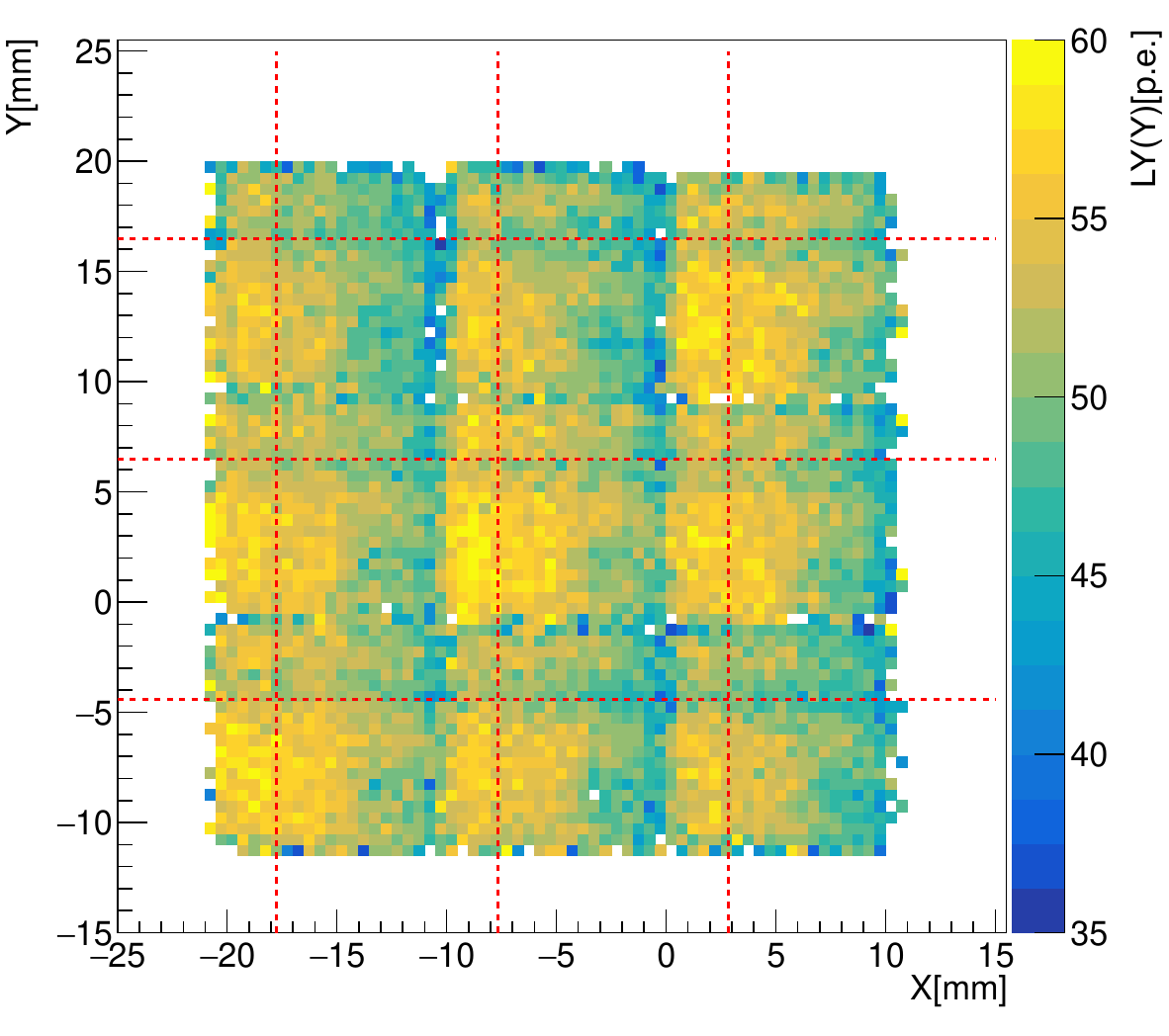}
&
\includegraphics[width=0.41\textwidth]{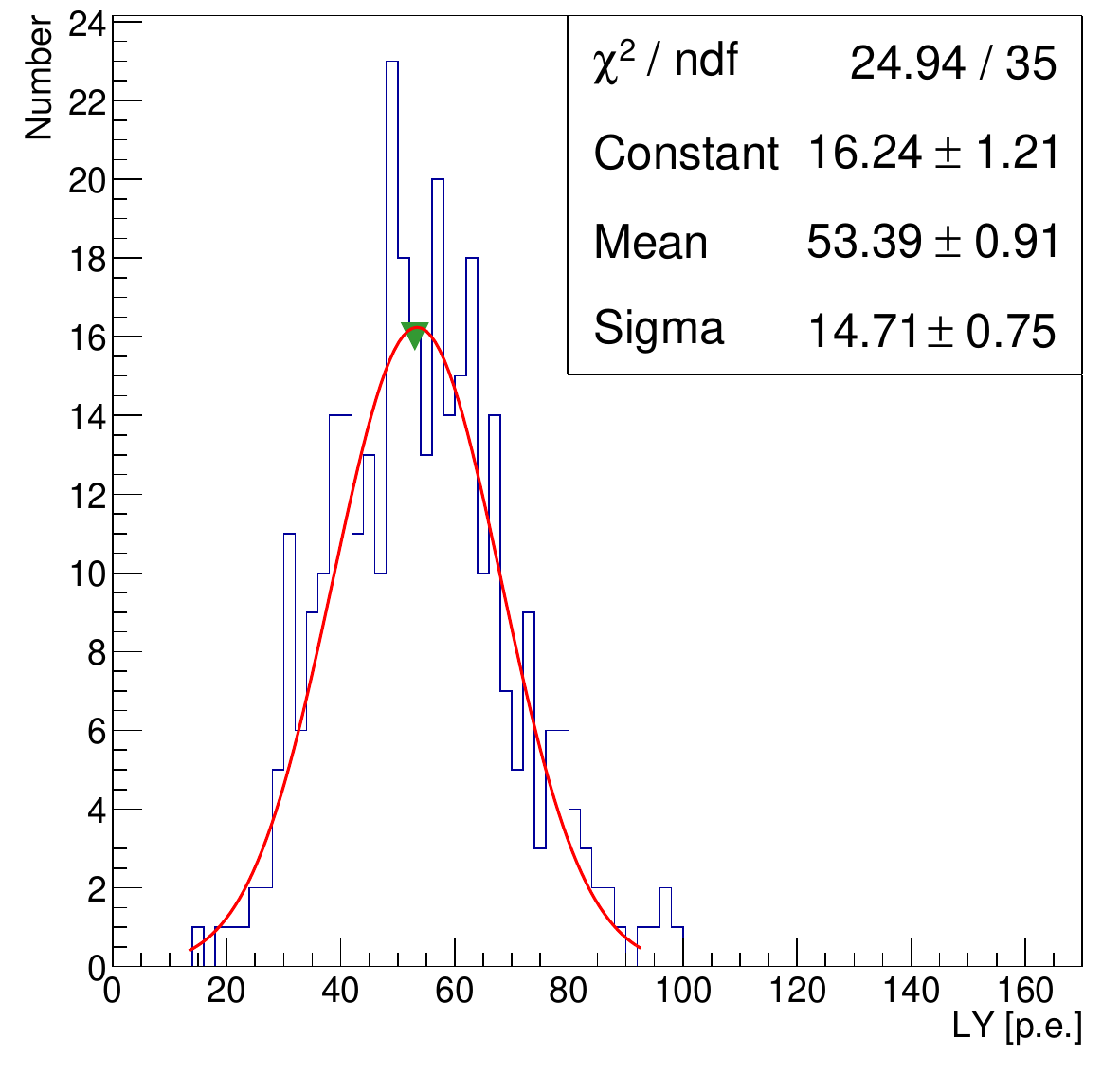}
\end{tabular}
\caption{Left: position-dependent light yield from $Y$-oriented fibers only. The red dotted lines indicate the positions of the $X$-oriented and $Y$-oriented fibers. Right: example of a Gaussian fit to the light yield distribution in a single bin. Green triangle marks the peak position identified using the \texttt{TSpectrum} algorithm.}
\label{fig:LY_Yonly}
\end{figure*}

Figures~\ref{fig:LY_Xonly} and \ref{fig:LY_Yonly} illustrate the light yield distributions obtained using signals from the $X$ and $Y$ fibers, respectively. In both cases, a spatial asymmetry is observed: the light yield increases as the particle track approaches the corresponding fiber axis. Specifically, the $X$-fiber map shows enhanced response near the horizontal ($X$) fiber, while the $Y$-fiber map exhibits higher signals near the vertical ($Y$) fiber. These observations are fully consistent with the physical geometry of the fibers embedded in the cubes and confirm that each fiber collects light most efficiently when the ionizing track is located in its immediate vicinity.

Examples of Gaussian fits to the light yield distributions in individual bins are shown on the right panels of each Figure. The use of fitted mean values helps to suppress statistical fluctuations and ensures reproducible behavior even in bins with limited statistics.

The analysis of light yield as a function of track position reveals non-uniformities within each scintillator cube at the sub-millimeter level. As expected, the signal increases near the embedded WLS fiber due to improved light collection efficiency. While the total light yield from both fibers shows relatively uniform behavior, individual contributions from $X$ and $Y$ fibers exhibit position-dependent gradients. These effects are important to consider in future reconstruction algorithms and simulation tuning, especially when evaluating energy resolution and response linearity at the cube level.
\label{sec:Cubes_LY}
\section{Average Light Yield Distribution Across Multiple Cubes}
To characterize the typical light collection behavior of a scintillator cube in the SuperFGD detector, we constructed an average LY map by combining position-dependent LY measurements from 27 individual cubes. This averaging procedure suppresses local variations due to imperfections in fiber alignment, or coupling efficiency, and reveals the general response pattern of a standard cube.

For each cube, the light yield was measured as a function of the particle track position with a bin size of 0.5~mm in both $X$ and $Y$ directions, as described in Section~\ref{sec:Cubes_LY}. The cube positions used for spatial alignment were determined in Section~\ref{sec:Cubes_position}. The LY distributions were then averaged bin-by-bin across all cubes. This was performed separately for three signal components:
\begin{itemize}
\item the combined signal from both $X$ and $Y$ fibers,
\item the signal from the $X$-fiber only,
\item the signal from the $Y$-fiber only.
\end{itemize}

For each bin, the LY distribution was fitted with a Gaussian function, and the mean of the fit was used as the representative value. This fitting-based approach ensures stability against statistical fluctuations and electronic noise.

\begin{figure*}
\centering
\begin{tabular}{ccc}
\includegraphics[width=0.33\textwidth]{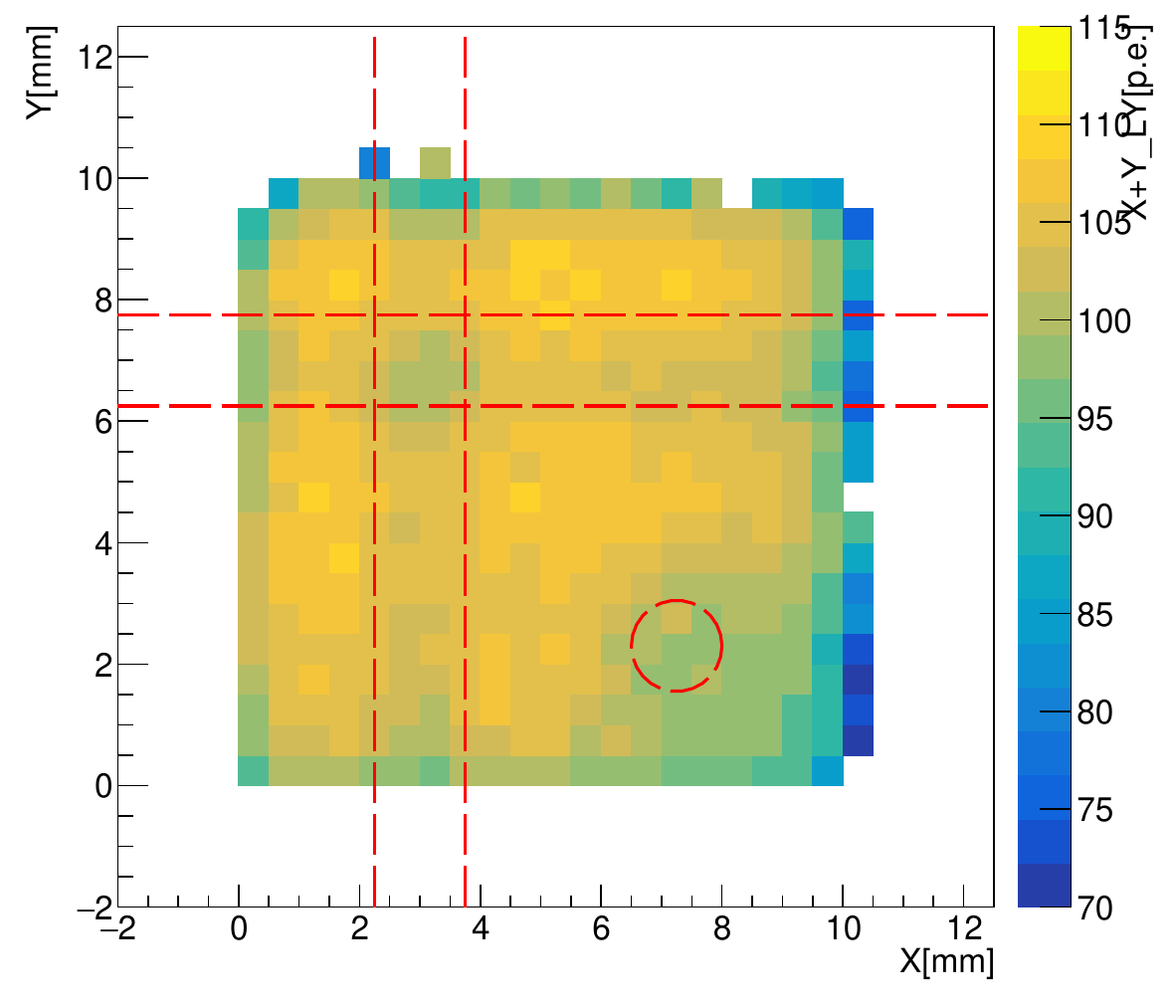} &
\includegraphics[width=0.33\textwidth]{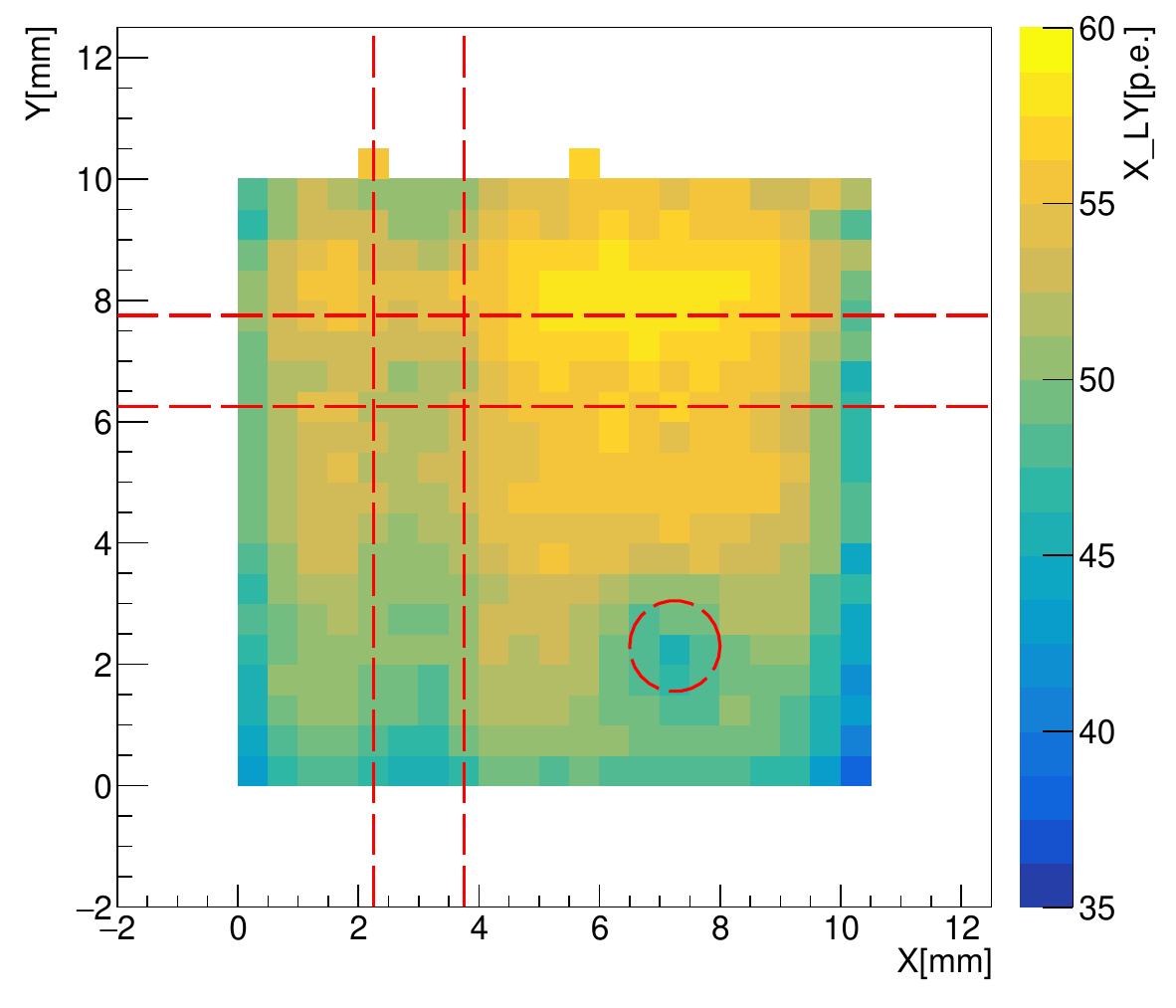} &
\includegraphics[width=0.33\textwidth]{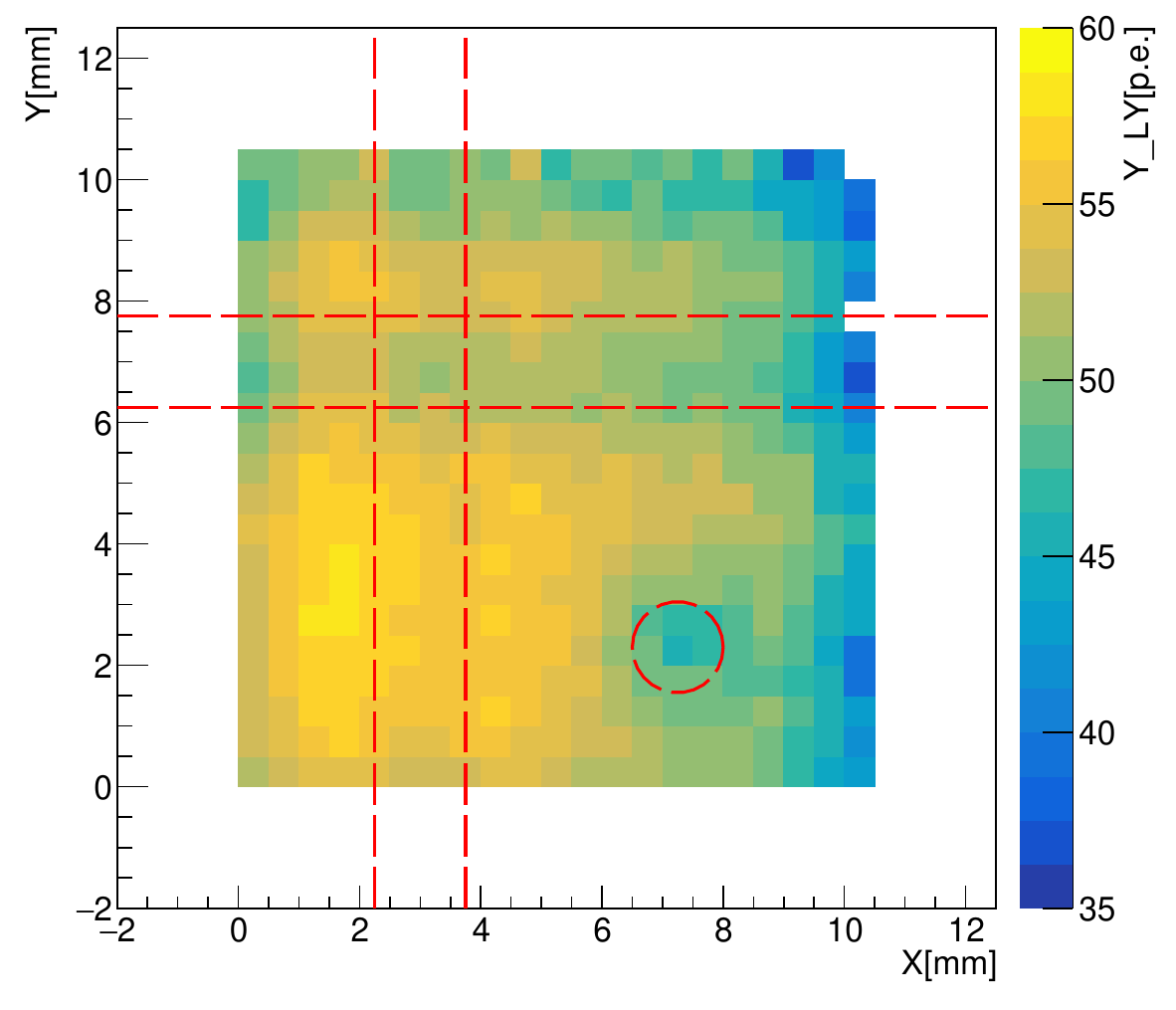}
\end{tabular}
\caption{Light yield maps  averaged over 27 cubes. Left: combined $X+Y$ fibers signals; middle: $X$ fiber only; right: $Y$ fiber only. Bin size is 0.5~mm. The features at $x \in [7,9]\text{~mm}, y \in [1,3]$~mm correspond to the presence of the $Z$-oriented fiber, leading to a localized reduction in light collection. The red dotted lines indicate the positions of the $X$-oriented and $Y$-oriented fibers. The red circle indicates the position of the $Z$-oriented fiber.}
\label{fig:Avr_LY_cube}
\end{figure*}

Figure~\ref{fig:Avr_LY_cube} presents the resulting average LY maps. The left panel shows the LY map obtained from the sum of $X$ and $Y$ fiber signals. The central zone of the cube exhibits relatively uniform response around 110 photoelectrons. A localized drop in LY to approximately 90 p.e. is observed in the region $x \in [7,9]$~mm, $y \in [1,3]$~mm, corresponding to the position of the $Z$-oriented fiber, which may reduce the light collection. Reduced LY is visible also near $x \in [2,4]$~mm and $y \in [6,8]$~mm, corresponding to the positions of the $Y$ and $X$ fibers, respectively.

The middle panel shows the contribution from the $X$ fibers only. The average LY is around 55 p.e., with a decrease to about 45 p.e. near the $Z$ fiber zone ($x \in [7,9]$, $y \in [1,3]$). As expected, increased LY is observed along the $Y$-axis at $y \in [6,8]$ where the $X$ fiber is closest to the track path. In contrast, a moderate drop to $\sim$48 p.e. appears near $x \in [2,4]$, corresponding to the $Y$ fiber.

The right panel shows the $Y$ fiber contribution. It exhibits the expected complementary pattern: higher LY values near $x \in [2,4]$ (i.e., closer to the $Y$ fiber) and lower values near $y \in [6,8]$, where the track is further away from the $Y$ fiber and closer to the $X$ fiber.

These averaged maps reflect the typical spatial dependence of light yield within a cube and are instrumental for detector simulation tuning, performance modeling, and reconstruction algorithm development in SuperFGD.
\label{sec:Avrage_LY}
\section{Consistency Check of Cube Alignment via Light Yield Maps}
To verify the geometric alignment and consistency of scintillator cubes in the $5\times5\times5$ SuperFGD prototype, we performed a comparative analysis of the spatial LY distributions. Specifically, we tested how well the LY maps of individual cubes match the average LY response under the assumption of ideal geometric alignment and uniform optical properties.

The average LY map was constructed as described in Section~\ref{sec:Avrage_LY}, by combining position-dependent LY distributions from 27 individually studied cubes. For each cube $k$, the measured light yield in a given bin $(i,j)$ is denoted by $LY_{ij}^{(k)}$. Deviations from the average map may occur due to statistical fluctuations or intrinsic optical differences between cubes.

A direct bin-by-bin comparison of individual LY maps is limited by low statistics in some regions. Therefore, to increase statistical power and focus on geometrical consistency, we compared 1D projections of the light yield along the horizontal ($X$) and vertical ($Y$) directions, separately for the $X$ and $Y$ readout fibers.

\begin{figure*}
\centering
\begin{tabular}{cc}
\includegraphics[width=0.47\textwidth]{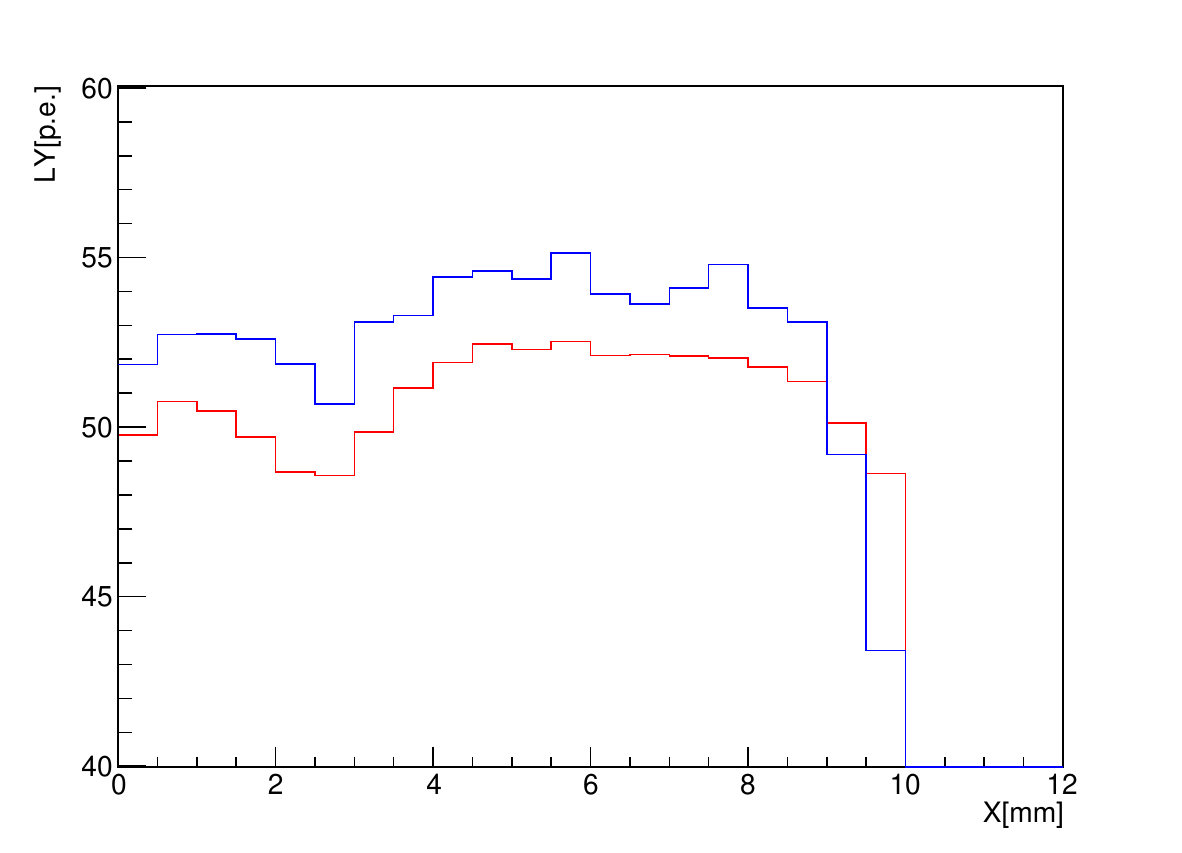} &
\includegraphics[width=0.47\textwidth]{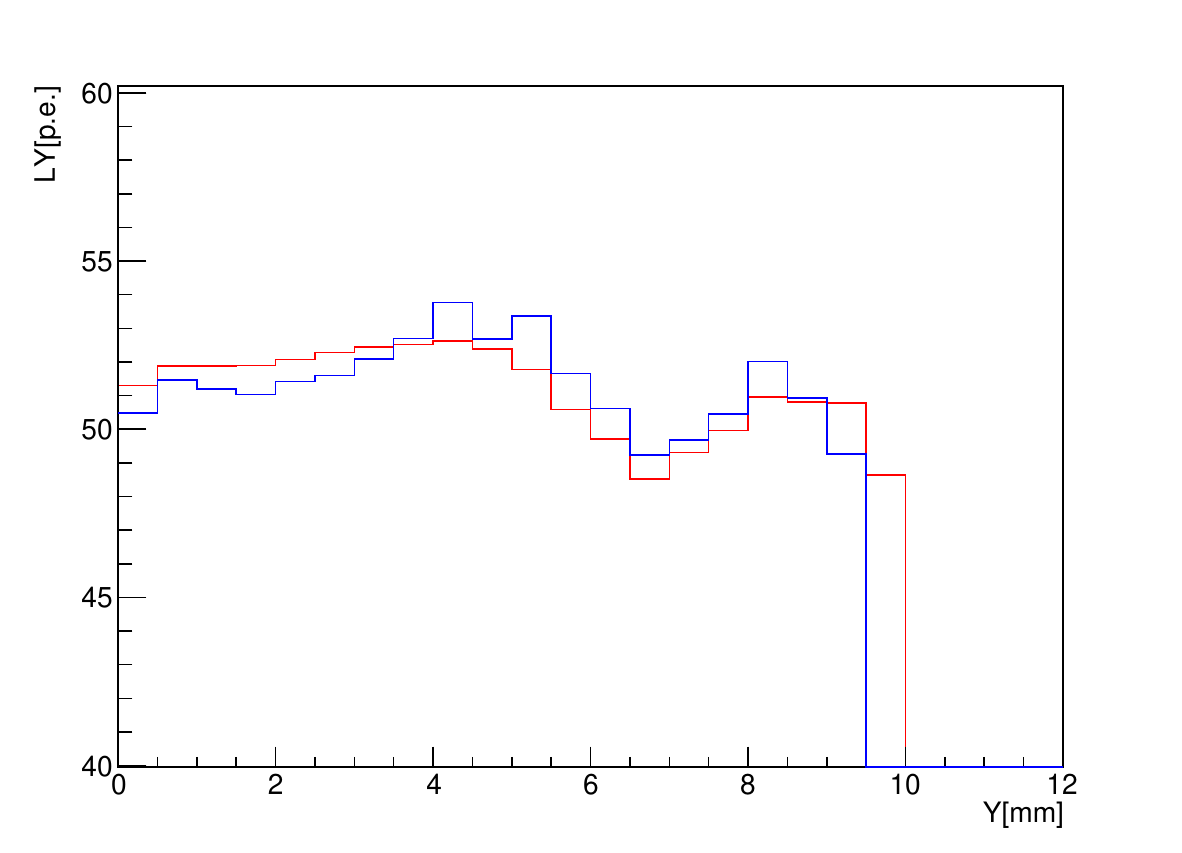}
\end{tabular}
\caption{1D projections of light yield from $X$ fibers along the $X$ direction (left) and from $Y$ fibers along the $Y$ direction (right). The red line shows the average over all cubes; blue lines show individual fiber responses of selected cubes.}
\label{fig:alignment}
\end{figure*}

Figure~\ref{fig:alignment} shows the results of this comparison:
\begin{itemize}
    \item Left panel: projection of light yield collected by the $X$ fibers along the $X$-axis;
    \item Right panel: projection of light yield from the $Y$ fibers along the $Y$-axis.
\end{itemize}

In both plots, the red line shows the average LY response over all cubes, while the blue lines represent individual fiber responses for selected cubes.

Of particular interest is the comparison at positions corresponding to the orthogonal fiber direction, i.e.:
\begin{itemize}
\item $X \in [2,4]$ for the $X$ fiber LY projection (where the $Y$ fiber is located),
\item $Y \in [6,8]$ for the $Y$ fiber LY projection (where the $X$ fiber is located).
\end{itemize}

These regions are sensitive to alignment mismatches, as they correspond to a local minimum in light collection due to proximity to the orthogonal fiber. Our study revealed that the individual and average LY curves closely overlap in these regions, with alignment deviations smaller than the bin size (0.5 mm) for all 27 cubes analyzed.

These results confirm the internal geometric consistency of the cube array and support the reliability of position-based LY averaging. This approach provides a sensitive probe of misalignments and validates both the mechanical assembly of the prototype and the assumptions used in detector simulation and analysis. It demonstrates that the mechanical precision of cube positioning is better than 0.5 mm across the detector volume.
\label{sec:alignment}
\section{Position-Dependent Optical Crosstalk Between Cubes}
An important characteristic of highly segmented scintillator detectors is the level of optical crosstalk between adjacent elements. In the SuperFGD, light produced in one cube can partially leak into neighboring cubes due to internal reflections, photon scattering or imperfections in optical isolation.  Such crosstalk may lead to signal smearing and affect the spatial resolution, and therefore must be carefully quantified and modeled.

In this section, we present a detailed study of the optical crosstalk as a function of the track position within a cube and the orientation of the readout fiber. The analysis is based on events where a charged pion crosses a well-identified cube (as defined in Section~\ref{sec:Cubes_position}), and signals in the four adjacent cubes are evaluated for correlated light leakage.

Crosstalk is evaluated individually for each of the four neighboring directions: left, right, up, and down. For every pion track passing through a cube, we examine the signal leakage into its adjacent neighbors sharing a face along the $X$ or $Y$ axis. In the following, we refer to these as “left”, “right”, “up”, and “down” neighbors. Directional contributions are isolated by selecting only one fiber signal in the neighboring cube. \textbf{Left and right} crosstalk (along the horizontal axis) and \textbf{up and down} crosstalk (along the vertical axis) are evaluated using the $Z$ fiber signal in the neighbor cubes. The analysis region was restricted to within 1.5~mm from the cube boundaries along $x$ to suppress contributions from events where the pion crossed the detector between cube rows.

\begin{figure*}
\centering
\begin{tabular}{cc}
\includegraphics[width=0.4\textwidth]{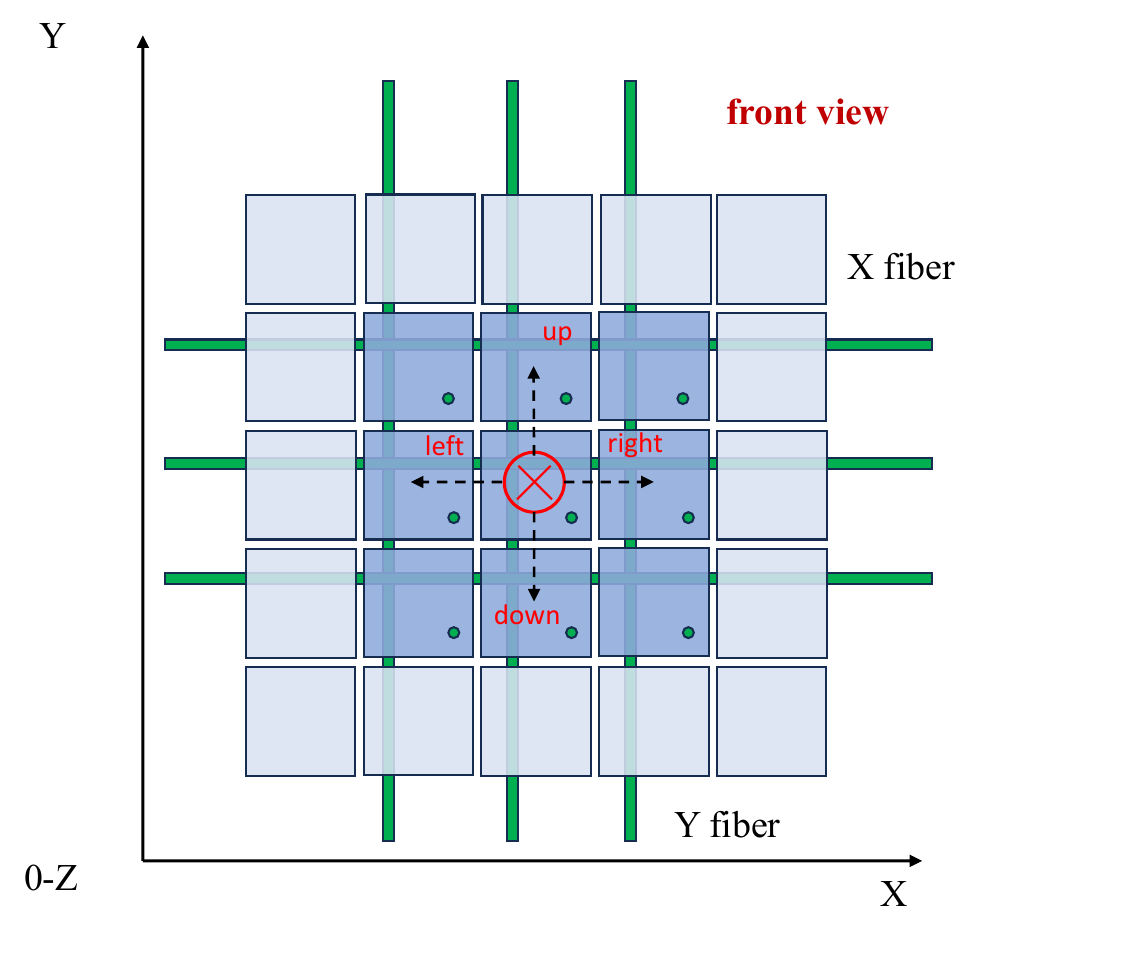} &
\includegraphics[width=0.55\textwidth]{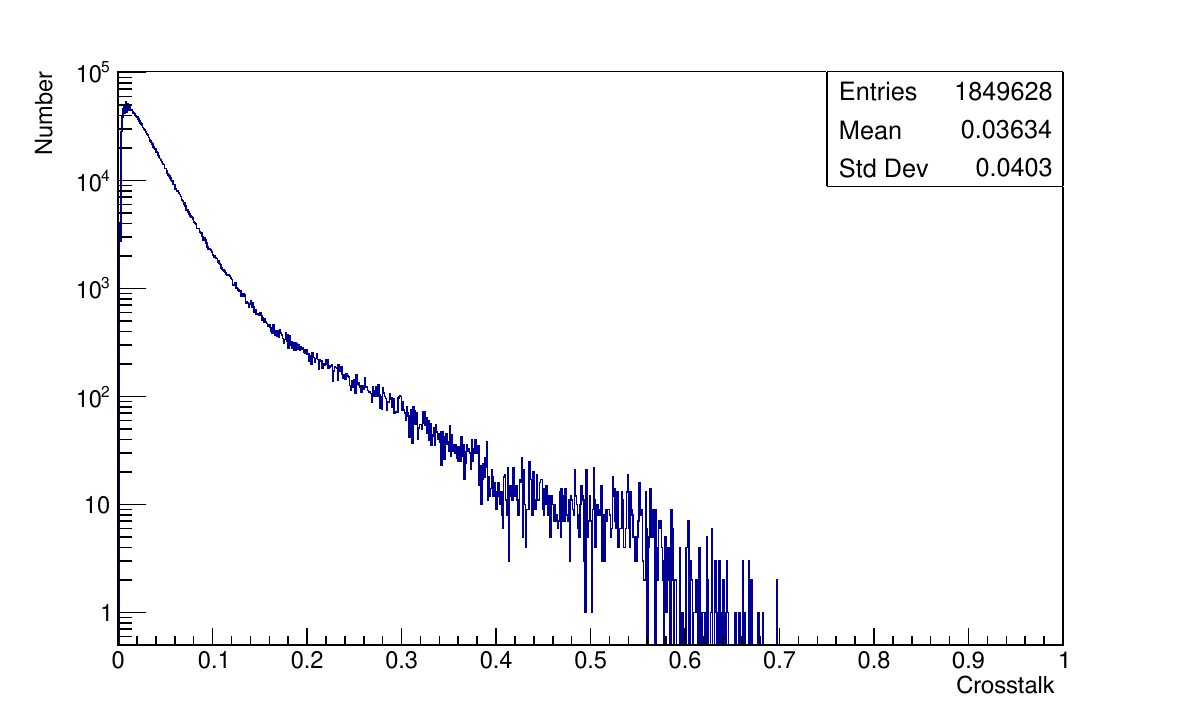}
\end{tabular}
\caption{Left: schematic illustration of crosstalk measurements. The red arrow shows the pion trajectory through the central cube; dashed arrows indicate the directions of potential signal leakage. Right: distribution of average crosstalk across $X$ and $Y$ directions.}
\label{fig:x-talk_scheme}
\end{figure*}

A schematic illustration is shown in Figure~\ref{fig:x-talk_scheme} (left). The red arrow denotes the pion track through the central cube, while the dashed arrows indicate possible light leakage directions. Crosstalk is calculated on an event-by-event basis as:

\begin{equation}
\label{eq:xtalk}
\text{crosstalk} = \frac{LY_{\text{Z fiber neighbor cube}}}{5 \times LY_{\text{avr}}},
\end{equation}

\noindent
where $LY_{\text{Z fiber neighbor cube}}$ is the light yield recorded in an adjacent cube, and $LY_{\text{avr}}$ is the average LY of the cubes from Figure~\ref{fig:different_LY} right. 

The distribution of the computed crosstalk values for all directions is shown in Figure~\ref{fig:x-talk_scheme} (right). The distribution mean is 3.63\%, consistent with earlier studies~\cite{Artikov2022}.

\subsection{Two-Dimensional Crosstalk Maps}

To explore how crosstalk depends on the track position within the cube, we constructed two-dimensional crosstalk maps for each pair of neighboring cubes. For each 0.5~mm bin in $(X,Y)$ coordinates, we selected events where the pion track passed through the cube and measured the signal in the corresponding adjacent cube. Crosstalk was computed as in Equation~\ref{eq:xtalk}, and the mean value from the distribution was used as the representative value for that bin.

This averaging approach suppresses statistical fluctuations and reveals systematic trends in light leakage. All maps are shown in the coordinate system of the main cube (crossing pion track), while the neighboring cube is positioned accordingly to the left, right, up or down.

\begin{figure*}
\centering
\begin{tabular}{cc}
\includegraphics[width=0.5\textwidth]{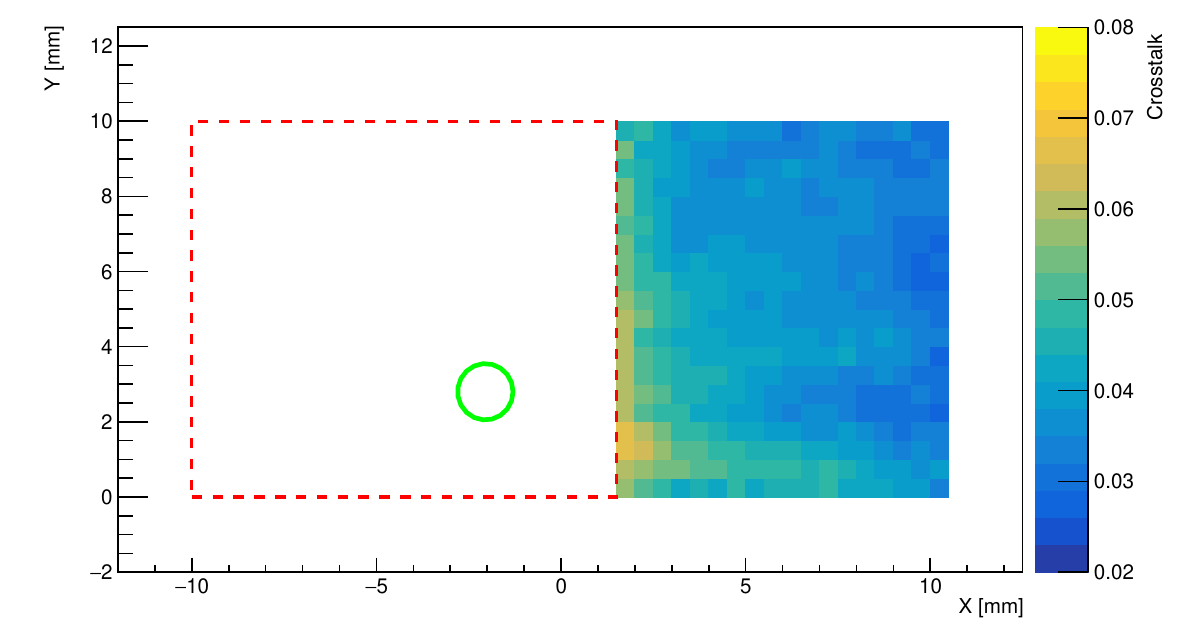} &
\includegraphics[width=0.5\textwidth]{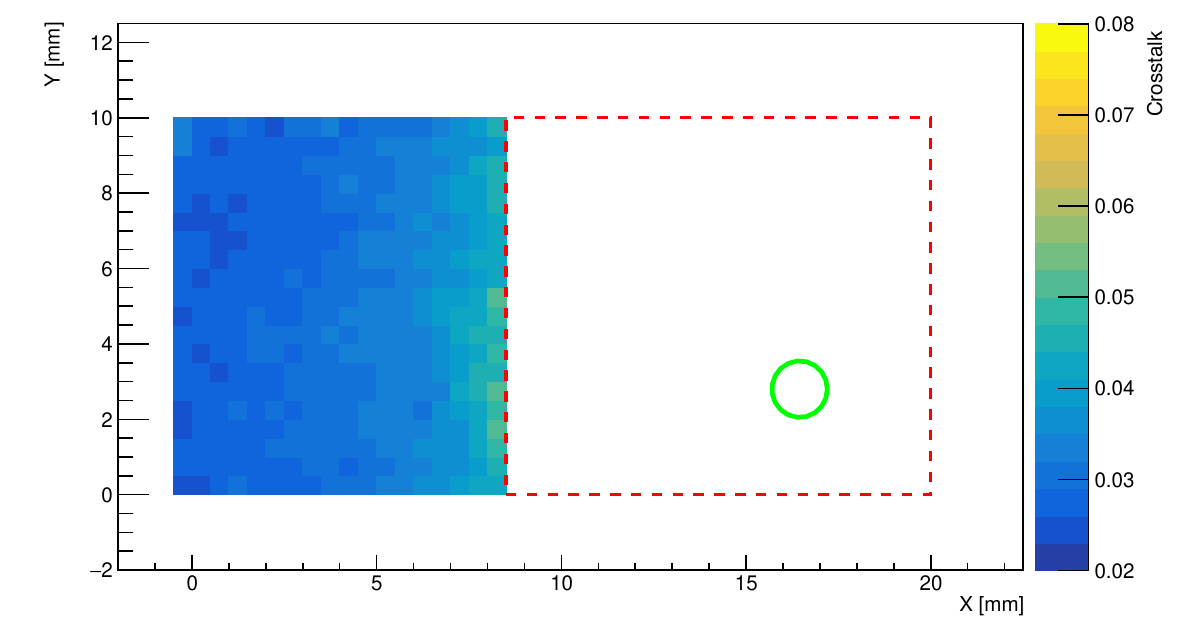}
\end{tabular}
\caption{2D maps of average crosstalk for right (right panel) and left (left panel) neighbor cubes. The color scale indicates mean of the distribution for each bin. The red dashed box indicates the projection of all five cubes sharing the same $Z$-oriented fiber in the neighboring column, from which the amplitude signal is used in Eq.~\ref{eq:xtalk}. The green circle indicates the position of $Z$-oriented fiber within the neighboring cubes.}
\label{fig:x-talk_2d_l&r}
\end{figure*}

Figure~\ref{fig:x-talk_2d_l&r} shows the 2D crosstalk maps for the \textbf{left} (left panel) and \textbf{right} (right panel) neighbors. In both cases, a clear enhancement of crosstalk is seen near the boundary with the main cube, where the pion track passes close to the shared face. For left and right neighbors, crosstalk values reach up to 6\% near the edge and fall below 2\% deeper into the cube. In Figure~\ref{fig:x-talk_2d_l&r} the red dashed lines show the projection of all five cubes sharing the same $Z$-oriented readout fiber. Since a single $Z$-fiber spans the full column of five cubes, its projection is wider than the individual cube boundary. This difference is due to a slight misalignment of the detector axis of approximately 1.5~mm over 5~cm with respect to the beam direction.

\begin{figure*}
\centering
\begin{tabular}{cc}
\includegraphics[width=0.47\textwidth]{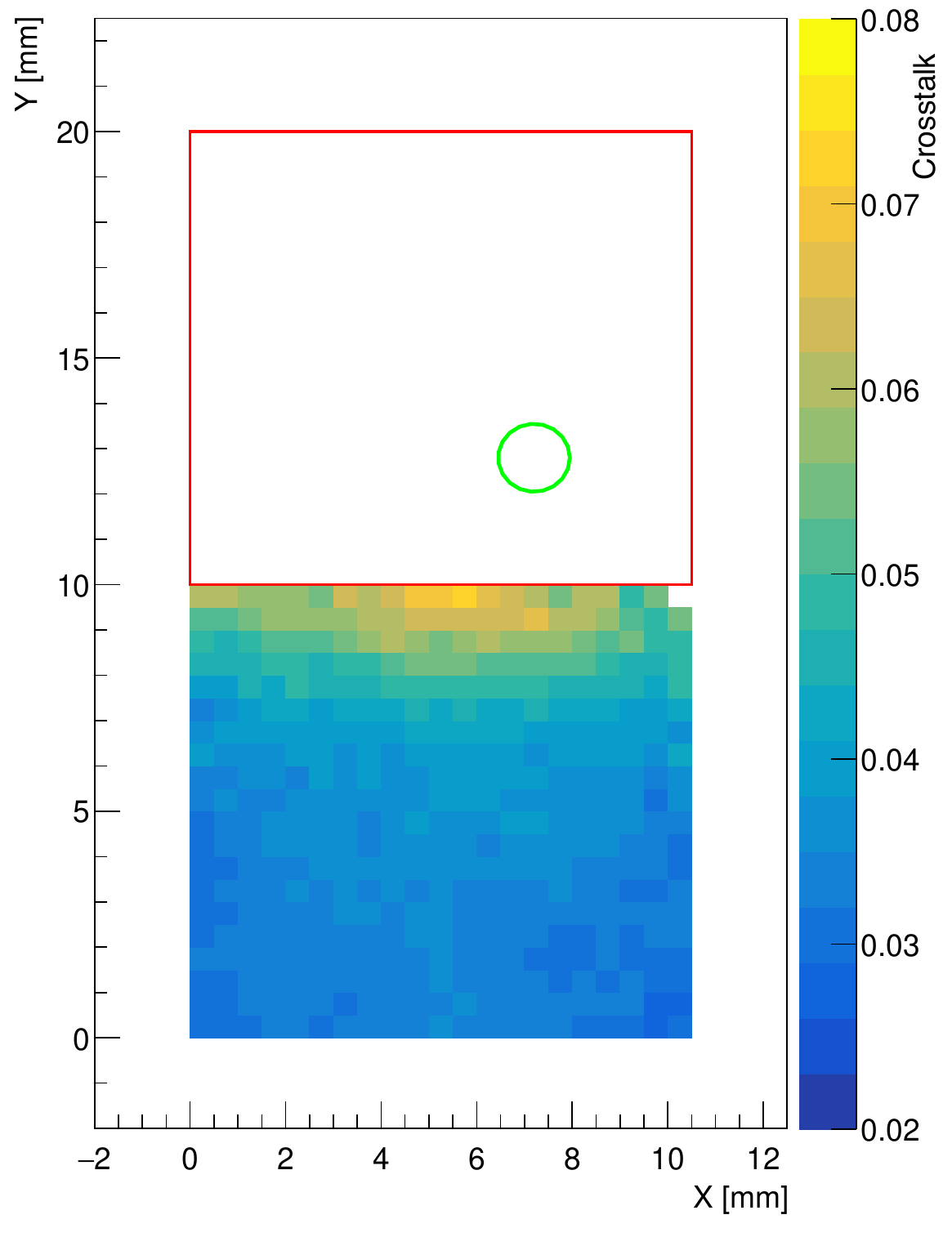} &
\includegraphics[width=0.47\textwidth]{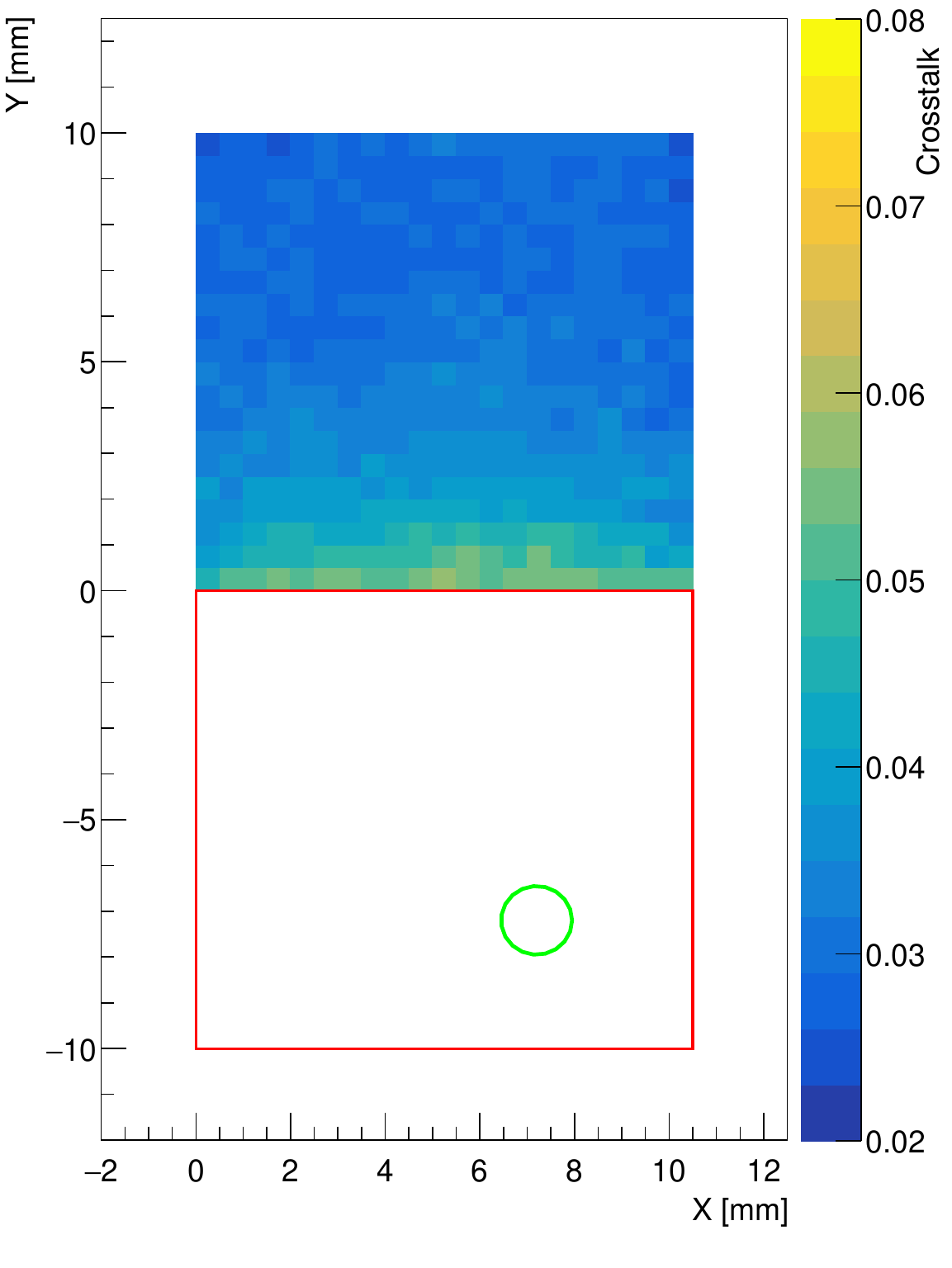}
\end{tabular}
\caption{2D maps of average crosstalk for up (left panel) and down (right panel) neighbor cubes. The red box indicates the boundaries of the neighboring cubes into which the crosstalk signal is measured. The green circle indicates the position of $Z$-oriented fiber within the neighboring cubes, from which the amplitude signal is used in Eq.~\ref{eq:xtalk}.}
\label{fig:x-talk_2d_u&d}
\end{figure*}

Similarly, Figure~\ref{fig:x-talk_2d_u&d} shows the crosstalk maps for the \textbf{up} (left panel) and \textbf{down} (right panel) neighbors. For up and down neighbors, crosstalk values reach up to 6\% near the edge and fall below 2\% deeper into the cube.

\subsection{One-Dimensional Crosstalk Maps}
\label{sec:xTalk1D}

To further clarify the spatial trends observed in the 2D crosstalk maps, we constructed one-dimensional (1D) crosstalk profiles as a function of the distance from the shared boundary between the cube hit by pion and each neighboring cube. This representation simplifies the interpretation by averaging over one coordinate direction, thus improving the statistical precision.

\begin{figure*}
\centering
\begin{tabular}{cc}
\includegraphics[width=0.47\textwidth]{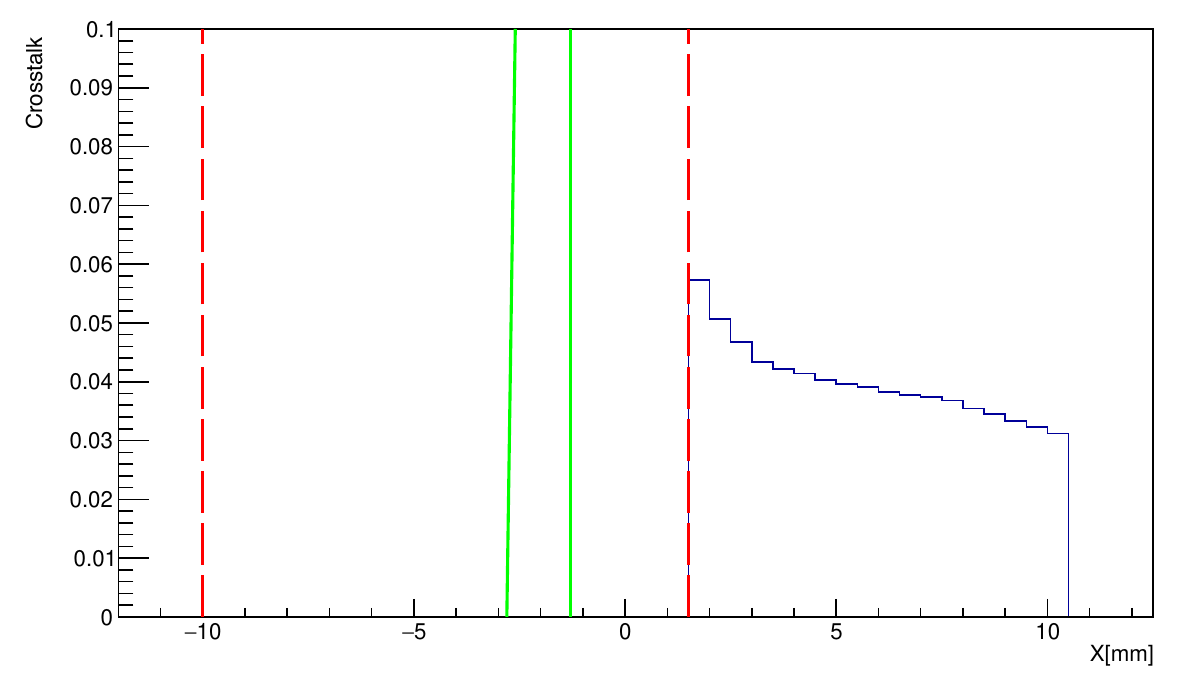} &
\includegraphics[width=0.47\textwidth]{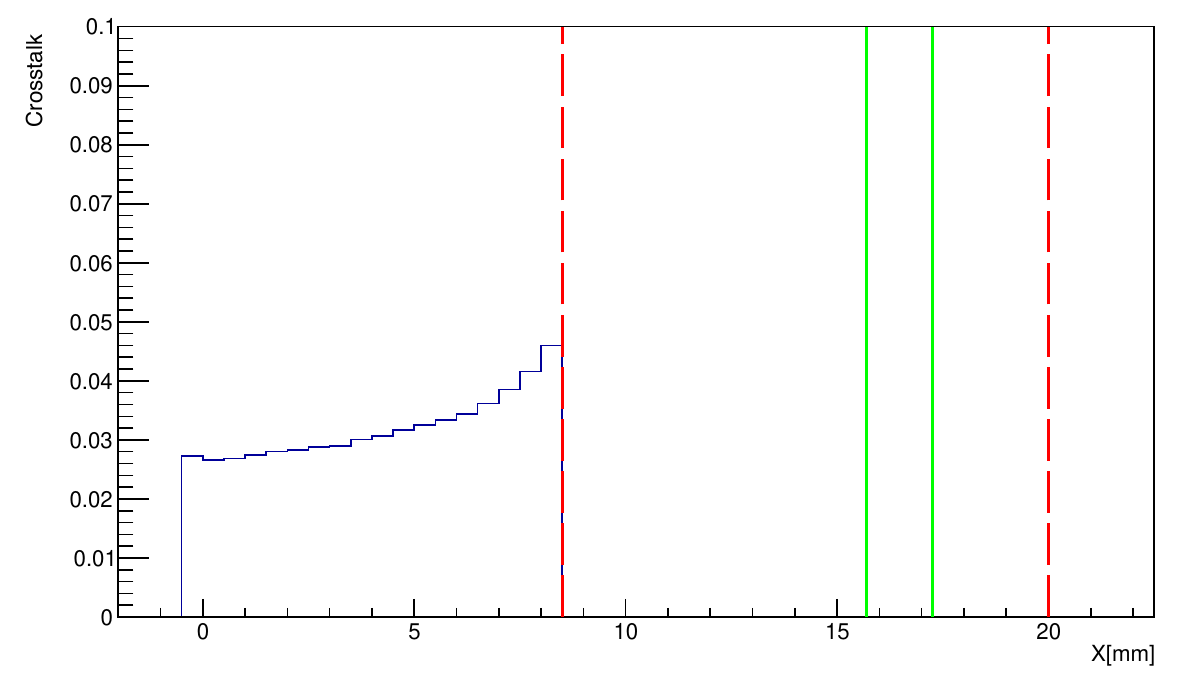}
\end{tabular}
\caption{One-dimensional crosstalk profiles for left (left panel) and right (right panel) neighbor cubes. Horizontal axis here indicates the distance from the cube boundary along $X$-axis. The red dashed lines indicate the projection of all five cubes sharing the same $Z$-oriented fiber in the neighboring column, from which the amplitude signal is used in Eq.~\ref{eq:xtalk}. The green solid lines indicate the position of $Z$-oriented fiber within the neighboring cubes.}
\label{fig:x-talk_1d_l&r}
\end{figure*}

\begin{figure*}
\centering
\begin{tabular}{cc}
\includegraphics[width=0.47\textwidth]{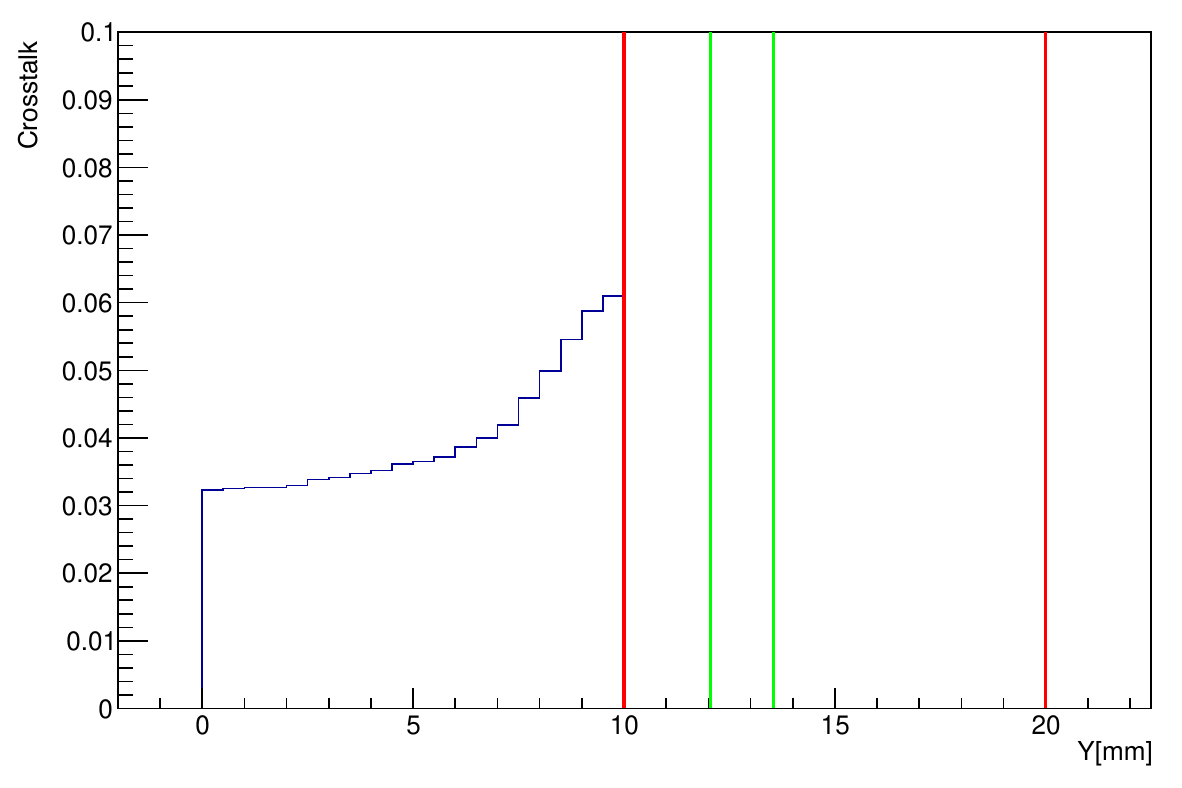} &
\includegraphics[width=0.47\textwidth]{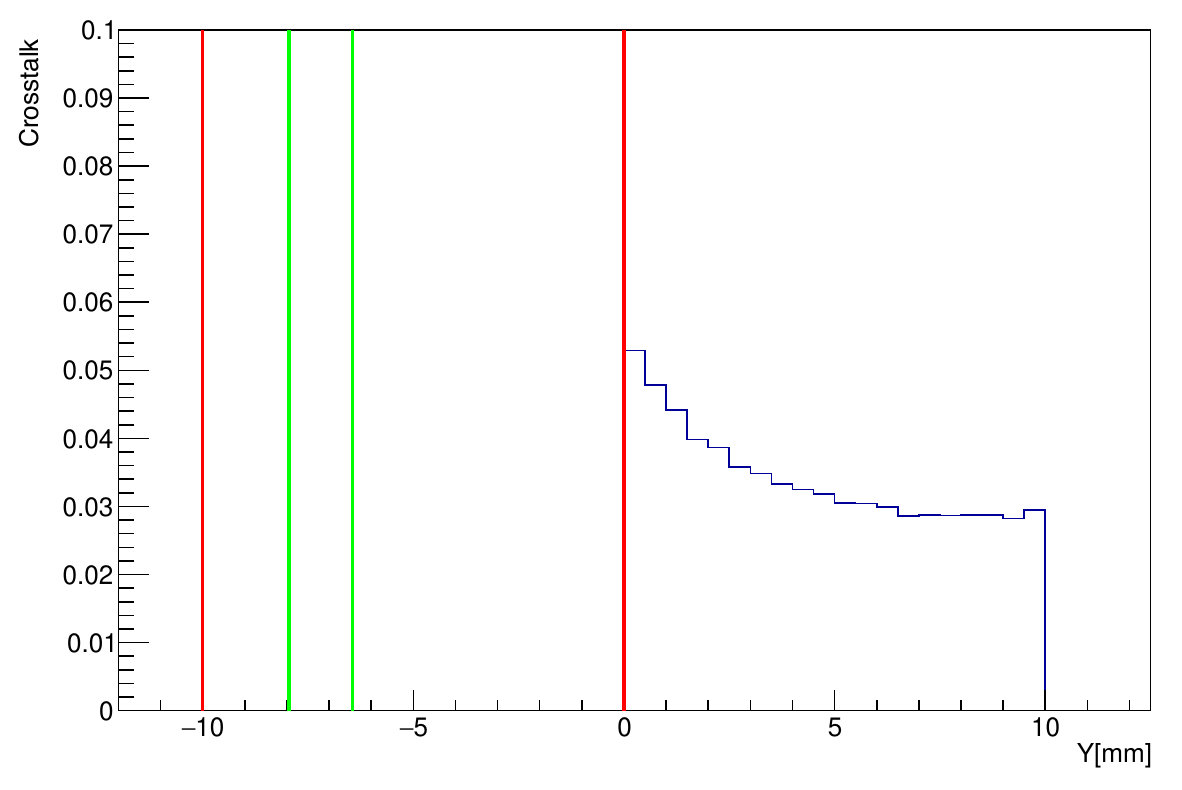}
\end{tabular}
\caption{One-dimensional crosstalk profiles for up (left panel) and down (right panel) neighbor cubes. The  horizontal axis here indicates the distance from the cube boundary along $Y$-axis. The red solid lines indicate the boundaries of the neighboring cubes into which the crosstalk signal is measured. The green lines indicate the position of $Z$-oriented fiber within the neighboring cubes, from which the amplitude signal is used in Eq.~\ref{eq:xtalk}.}
\label{fig:x-talk_1d_u&d}
\end{figure*}

Figures~\ref{fig:x-talk_1d_l&r} and \ref{fig:x-talk_1d_u&d} show the crosstalk behavior in the \textbf{left}, \textbf{right}, \textbf{up}, and \textbf{down} directions, respectively. For each 0.5~mm bin from the shared interface, we selected events where the pion track passed through the cube and evaluated the signal in the corresponding adjacent cube. As in the 2D case, the crosstalk value in each bin was computed using Equation~\ref{eq:xtalk}, and the resulting distribution mean was used as the representative crosstalk value. In Figures~\ref{fig:x-talk_1d_l&r} the red dashed lines show the projected extent of all five cubes sharing the same $Z$-oriented readout fiber. Since a single $Z$-fiber spans the full column of five cubes, its projection is wider than the individual cube boundary. This difference is due to a slight misalignment of the detector axis of approximately 1.5~mm over 5~cm with respect to the beam direction.

In the Figure~\ref{fig:x-talk_1d_l&r}, the crosstalk to the left and right neighbor decreases gradually from approximately 6\% to 3\% as the track moves away from the boundary. The average value of crosstalk to the left cube is slightly higher that to the right cube due to increased distance to the $Z$ fiber inside neighbor cube. 

Similarly, in Figure~\ref{fig:x-talk_1d_u&d}, the up and down crosstalk direction shows a decreasing trend from about 6\% to 3\% as the track moves away from the boundary. The average value of crosstalk to the upstream cube is a little bit higher than that to the downstream cube due to the increased distance to the $Z$ fiber inside the neighboring cube.  

These 1D profiles complement the 2D maps presented in Figures~\ref{fig:x-talk_2d_l&r} and \ref{fig:x-talk_2d_u&d}. The key difference is that each bin in the 1D profiles contains roughly 10 times more events than a bin in the corresponding 2D maps, resulting in smoother and statistically more robust behavior.

Together, the 1D and 2D maps provide a comprehensive view of the spatial behavior of optical crosstalk in the SuperFGD. These results are valuable inputs for detector simulation, calibration, and potential correction algorithms to improve spatial resolution and hit assignment.
\label{sec:Cross-talk}
\section{Monte Carlo Simulation and Comparison with Data}
To complement the experimental measurements and support the interpretation of the detector response, a dedicated Monte Carlo simulation of the SuperFGD prototype was developed. The simulation reproduces the geometry, materials, and basic optical properties of the $5\times5\times5$ scintillator cube array tested in the beam, including the response of the proportional chambers.

The primary goals of the simulation are:
\begin{itemize}
\item to model the spatial distribution of LY within individual cubes and compare it with experimental data;
\item to estimate the contribution of optical crosstalk between neighboring cubes and validate the observed position-dependent behavior;
\item to provide realistic input for detector reconstruction, calibration, and performance studies.
\end{itemize}

\subsection{Simulation Framework and Detector Model}
The simulation was implemented in \texttt{Python} with photon-level tracking. The detector model consisted of a $5\times5\times5$ cube array, with the central $3\times3\times3$ instrumented cubes representing the region analyzed in the beam test. The geometry included individual scintillator cubes, WLS fibers embedded along the three orthogonal axes, and reflective surfaces representing the diffuse reflector layer. Photon interactions such as scintillation, absorption, re-emission, and boundary reflections were fully simulated.

\begin{figure*}
\centering
\includegraphics[width=0.95\textwidth]{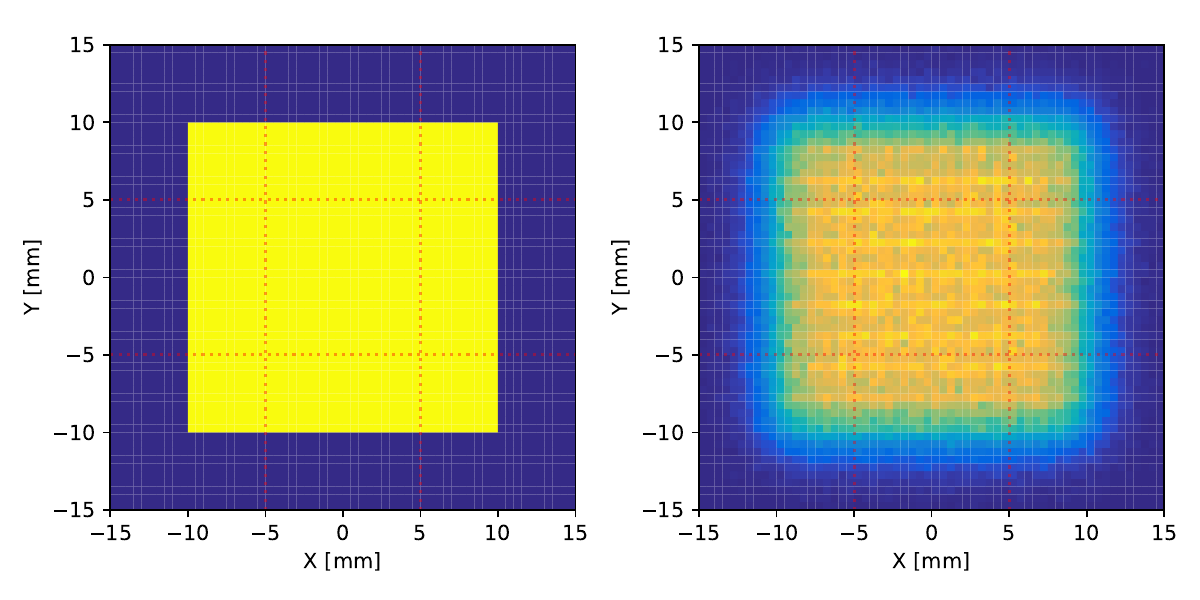}
\caption{Track distribution before (left) and after (right) applying proportional chamber reconstruction response.}
\label{fig:mc_proportional_chamber_systematics}
\end{figure*}

Charged pion tracks were generated with angular distributions consistent with experimental data. The proportional chamber system was modeled to reproduce track reconstruction uncertainties. Figure~\ref{fig:mc_proportional_chamber_systematics} shows the track of pion distribution in the central area $x\in[-10; 10]\ \text{mm},\ y\in[-10;10]\ \text{mm}$ before and after applying proportional chamber response. Energy depositions along the straight track segments in a cube followed a Landau distribution to represent ionization fluctuations.

Light production was modeled in two media:
\begin{itemize}
\item scintillator: $10{,}000$ photons/MeV,
\item WLS fiber material: $600$ photons/MeV (estimated from auxiliary calibration measurements).
\end{itemize}

No photons were generated in the air gaps or outside the scintillator. A $60~\mu$m thick layer represented the chemically etched reflective coating, treated as a region with zero light production.

\begin{figure*}
\centering
\includegraphics[width=0.95\textwidth]{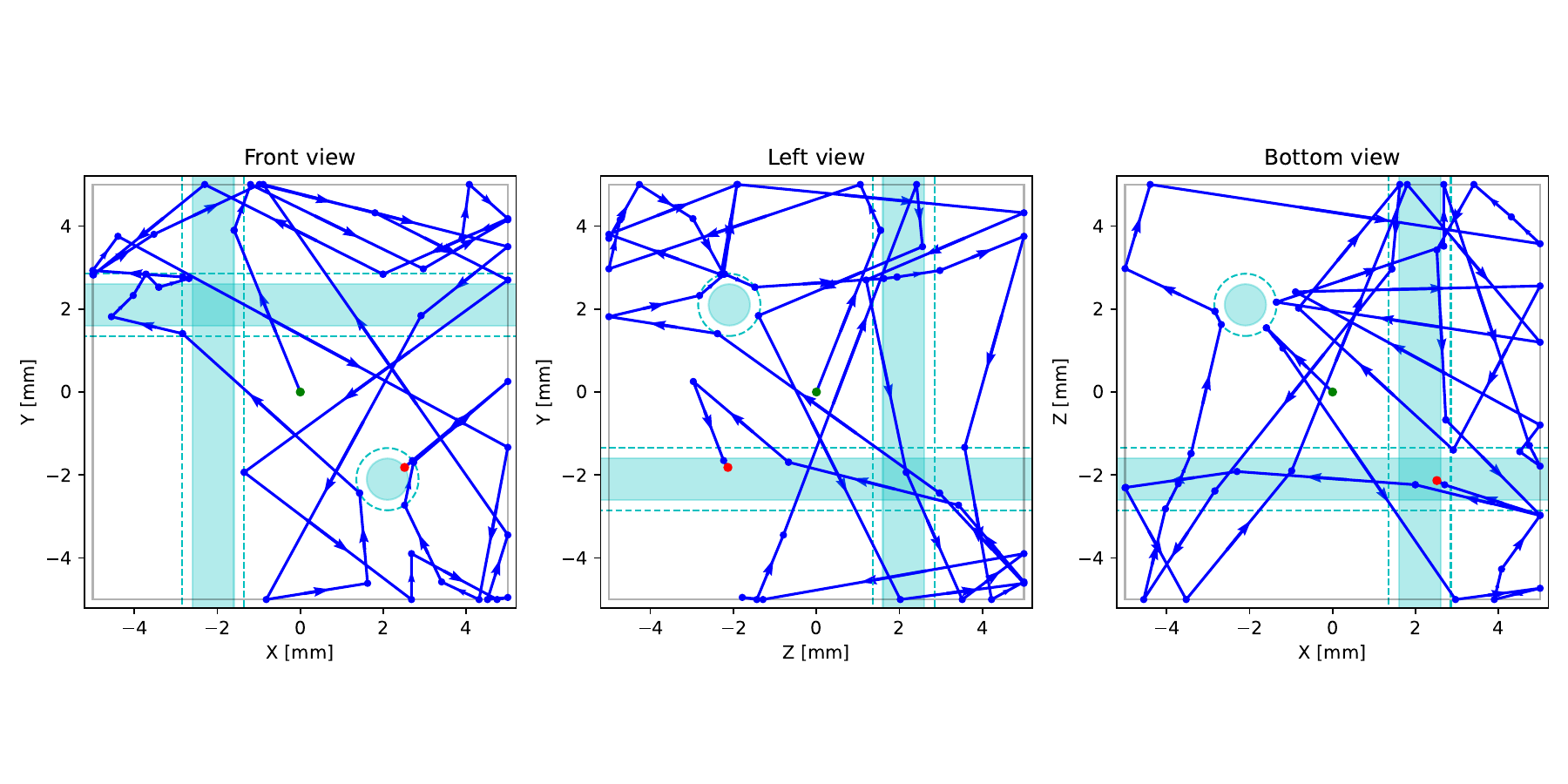}
\caption{Example of photon propagation inside a cube.}
\label{fig:mc_photon_propogation}
\end{figure*}

Photon emission was isotropic. Spectral effects and polarization were not included. Each photon propagated until absorption, with an exponential attenuation length of 100 mm. During propagation, photons could undergo reflection or refraction at cube walls, holes, or fiber surfaces. A photon was counted as “captured” if it entered the fiber volume. An example of photon propagation inside a cube is shown in Figure~\ref{fig:mc_photon_propogation}.

Attenuation along the WLS fiber was not included in the simulation. In the experimental setup, all instrumented cubes were located at the same distance from the SiPM ($35 \pm 1$~cm), so fiber attenuation affects all channels uniformly and is effectively absorbed into the per-channel normalization procedure described in Section~8. The residual variation of $\pm$1~cm between cube positions leads to a negligible difference in attenuation between channels.

\subsection{Optical Properties}
Surface interactions were modeled using the GLISUR micro-facet model. In this approach, surface roughness is parameterized by a “polish” value in the range [0,1].
\begin{itemize}
\item Cube walls were modeled with diffuse reflections ($\text{polish}=0.0$), with a constant reflectivity of 0.96.
\item Hole and fiber surfaces were modeled as polished ($\text{polish}=1.0$).
\end{itemize}

The scintillator refractive index was set to 1.59, while the outer fiber cladding was set to 1.49. Reflection probabilities at interfaces were calculated using Fresnel equations averaged over photon polarization.
\subsection{SiPM Response}

The SiPM response was not explicitly simulated. Instead, photons captured by fibers were converted into p.e. equivalents using a global calibration coefficient. This coefficient provided correspondence between MC photon counts (“MC units”) and experimentally measured p.e. It was tuned by matching the average LY obtained in simulation with the reference value from calibration data (see Section~\ref{sec:Calibration}).

Global calibration coefficient was acquired via a pixel-binned fit of the experiment m.p.v. pixel map by the MC m.p.v. map. Areas of the fibers were excluded from the fit. The fiber attenuation and PDE were not modeled exclusively and were included in the global coefficient.

\subsection{Light Yield Maps: Simulation vs Data}

\begin{figure*}
\centering
\includegraphics[width=0.95\textwidth]{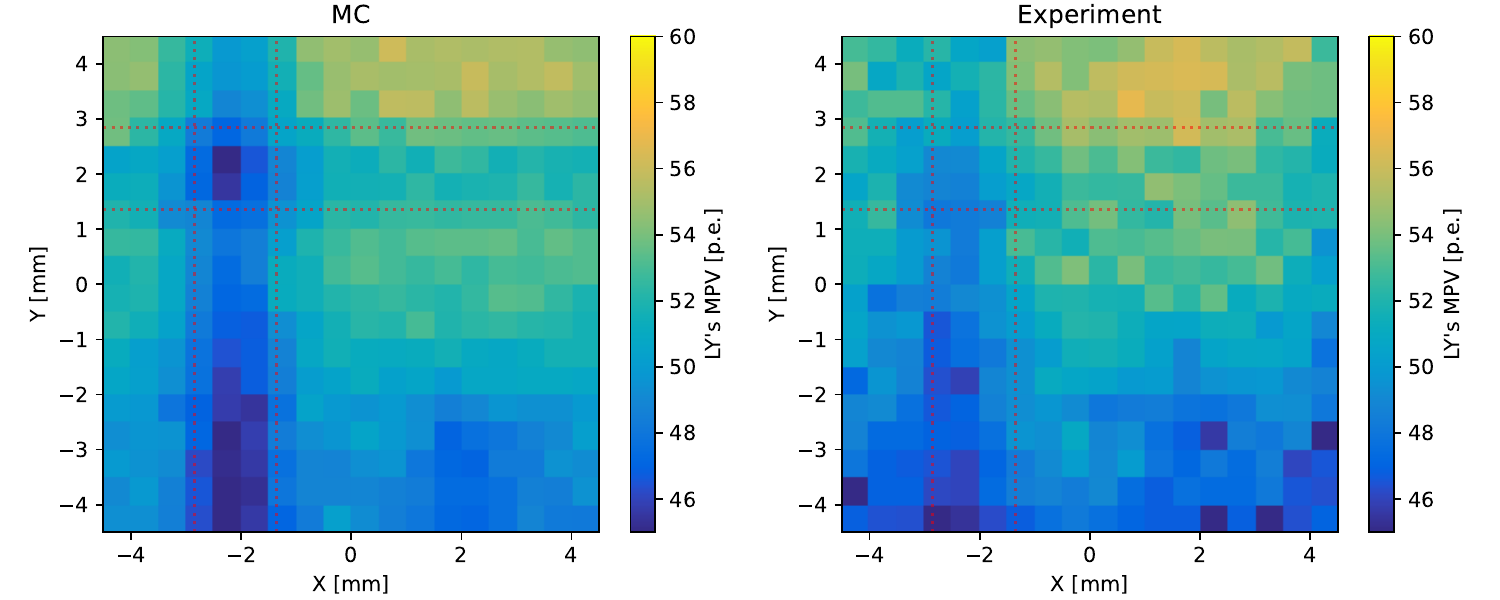}
\caption{Comparison of light yield maps for the $X$-fiber: (left) simulated LY for the central cube, (right) experimental average LY.}
\label{fig:mc_vs_data_ly}
\end{figure*}

To validate the model, we compared the simulated LY distribution with the average map obtained from experimental data (Section~\ref{sec:Avrage_LY}). For simplicity, the comparison was restricted to the central cube in simulation and the average over all cubes in data. The analysis region was limited to $x\in[-4.5;4.5]\ \text{mm},\ y\in[-4.5;4.5]\ \text{mm}$ in order to suppress edge effects. This choice is motivated by the alignment studies presented in Section~\ref{sec:alignment}, which showed that cube positions are matched within 0.5 mm. The resulting comparison is shown in Figure~\ref{fig:mc_vs_data_ly}.

The positions of the $Z$-fiber and $Y$-fiber are visible in the $X$-fiber LY maps of Figure~\ref{fig:mc_vs_data_ly}. However, in experimental data the fiber structures appear significantly smoother compared to the MC. This difference can be attributed to small ($<0.5$ mm) misalignments between cubes in the experimental setup, which smear out sharp features that remain visible in the idealized MC geometry. 

To further illustrate the observed differences, one-dimensional (1D) LY profiles were constructed by slicing the cube into three sections along each coordinate axis. The $X$-axis slices correspond to the bottom, middle, and top regions of the cube, while the $Y$-axis slices correspond to the left, middle, and right regions. This approach provides a clearer view of local LY variations and enhances statistical precision.

\begin{figure*}
\centering
\includegraphics[width=0.95\textwidth]{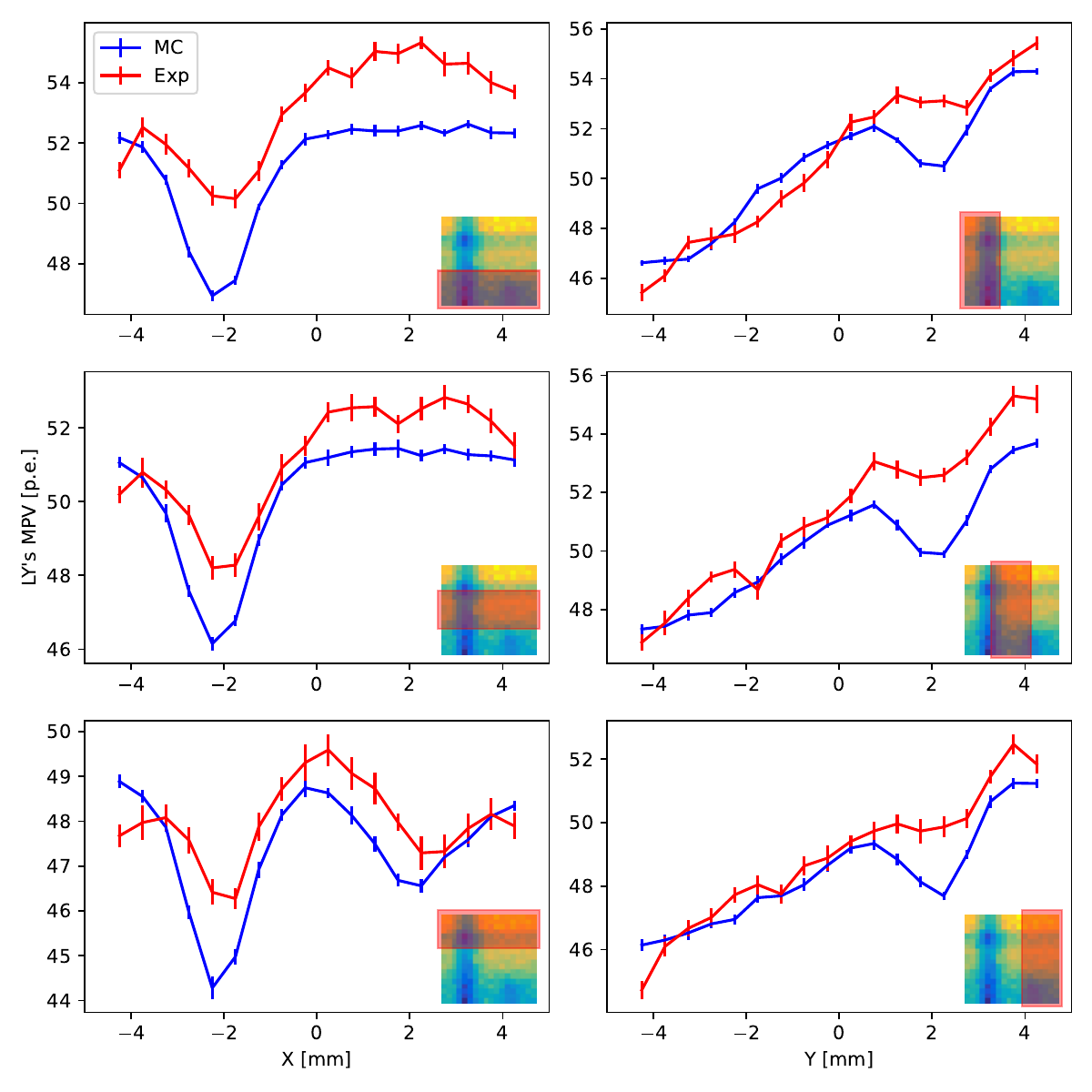}
\caption{Comparison of one-dimensional LY profiles between data (red) and MC simulation (blue). Left column: profiles along the $X$-axis for the bottom, middle, and top thirds of the cube. Right column: profiles along the $Y$-axis for the left, middle, and right thirds of the cube. The small 2D maps shown in the lower right corners of each panel indicate the corresponding region of the LY map used to extract the 1D profile.}
\label{fig:slise_LY_map}
\end{figure*}

Figure~\ref{fig:slise_LY_map} (right) shows the LY profiles along the $Y$-axis for the left, middle, and right thirds of the cube. In the left region, the $X$-fiber structure is clearly seen in the MC simulation (blue line), whereas in the data (red line) the corresponding enhancement appears smoother and less localized. In the right region, the $Z$-fiber feature is well reproduced in both data and MC, while the $X$-fiber signature remains more diffuse in data, consistent with small misalignment between cubes.

Similarly, Figure~\ref{fig:slise_LY_map} (left) presents the LY profiles along the $x$-axis for the bottom, middle, and top thirds of the cube. The positions of the $Y$- and $Z$-fibers are clearly visible in both MC and data. However, the effect of reduced LY in fiber regions (the effect of the reduced active material in the fiber region) is more pronounced in the simulation, reflecting the idealized data assumed in the model. In contrast, the data exhibit smoother transitions presumably due to minor misalignment between cubes.

These comparisons demonstrate that the simplified MC model successfully reproduces the main features of the measured light yield distribution, including the locations of WLS fibers and the reduced response in their vicinity. The smoother appearance of the experimental data indicates that small mechanical shifts, reflective layer imperfections, and diffuse photon transport contribute to the realistic light distribution observed in the prototype.

\subsection{Crosstalk Maps: Simulation vs Data}
A similar procedure was applied to evaluate the optical crosstalk using the Monte Carlo simulation. Crosstalk maps were constructed from MC using the same definition as in Equation~\ref{eq:xtalk} and directly compared to the experimental measurements presented in Section~\ref{sec:xTalk1D}. The comparison highlights both the common features (such as enhancement near cube boundaries and a rapid falloff deeper into the volume) and the differences that may arise from simplified assumptions in the optical model (e.g., omission of wavelength dependence and detailed SiPM response).

\begin{figure*}
\centering
\includegraphics[width=0.95\textwidth]{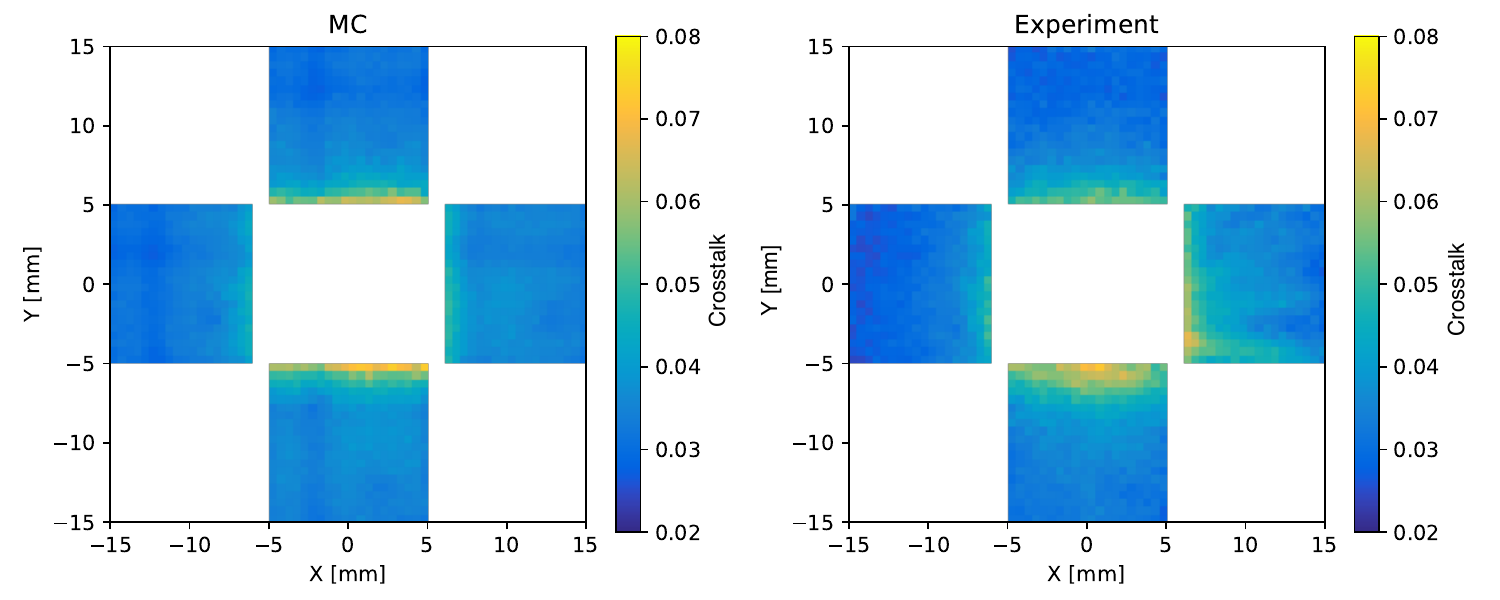}
\caption{Comparison of two-dimensional crosstalk maps between data and Monte Carlo simulation. Both show enhanced light leakage near the cube boundaries and a rapid decrease toward the cube center.}
\label{fig:MC_x-talk_2D}
\end{figure*}

The comparison between simulated and measured 2D crosstalk maps is shown in Figure~\ref{fig:MC_x-talk_2D}. The analysis region was restricted to within 1.5~mm from the cube boundaries along $x$ to suppress contributions from events where the pion crossed between cube rows.

\begin{figure*}
\centering
\includegraphics[width=0.95\textwidth]{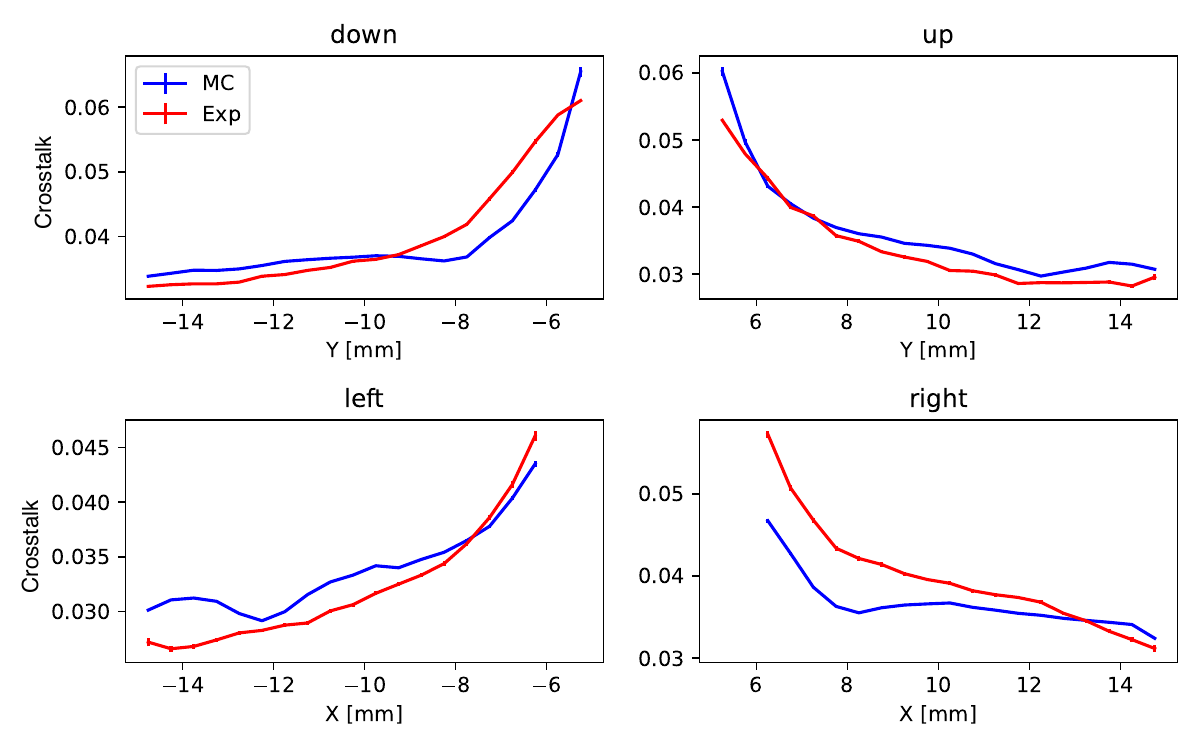}
\caption{Comparison of one-dimensional crosstalk profiles for down (top left), up (top right), left (bottom left), and right (bottom right) neighboring cubes. The $X$-axis indicates the distance from the cube boundary. Experimental data (red) and Monte Carlo simulation (blue).}
\label{fig:MC_x-talk_1D}
\end{figure*}

To further illustrate the agreement, one-dimensional (1D) crosstalk profiles were constructed along each coordinate axis, following the same binning and averaging procedure as in Section~\ref{sec:xTalk1D}. The results, shown in Figure~\ref{fig:MC_x-talk_1D}, demonstrate overall consistency between data and simulation. The main trends -- peak crosstalk near the boundaries (5–7\%) and smooth falloff to below 3\% inside the cube -- are well reproduced. Minor deviations can be attributed to cube misalignment, reflective surface imperfections, and simplified optical treatment in the model.

Overall, the simulation reproduces the main qualitative features of the optical crosstalk observed in the data, confirming that the implemented optical model captures the dominant light transport mechanisms. The remaining discrepancies point to the importance of including spectral dependencies, detailed SiPM response, and realistic surface imperfections in future iterations of the simulation to achieve full quantitative agreement.
\label{sec:mc_simulation}
\section{Conclusion}
In this study, we presented a detailed characterization of a $5\times5\times5$ cubes prototype of the SuperFGD detector using a pion beam at the SC-1000 synchrocyclotron. The prototype, equipped with wavelength-shifting fibers and SiPM readout, was tested with high spatial resolution to evaluate key detector performance parameters.

The average light yield per fiber was measured to be approximately 51~p.e. for minimum ionizing particles, with up to 100~p.e. collected from two orthogonal fibers per cube.

Position-resolved measurements within individual cubes, with a bin size of 0.5~mm, revealed the spatial non-uniformity of light collection. The combined response from two orthogonal fibers showed good homogeneity across most of the cube volume, with localized variations near the embedded fiber axes. Averaging over 27 cubes yielded a representative light response map, useful for simulation tuning and reconstruction optimization.

Optical crosstalk between adjacent cubes was studied in four directions (left, right, up, and down). The measured crosstalk values ranged from 2\% to 6\%, depending on the particle track position relative to cube boundaries. Detailed two-dimensional and one-dimensional crosstalk maps were produced, revealing spatial dependence and asymmetries between directions.

To complement the experimental measurements, a dedicated Monte Carlo simulation of the prototype was developed, reproducing the detector geometry and photon transport. The MC results successfully reproduced the main features of the measured light yield distributions and the position-dependent crosstalk behavior, including enhanced signal near cube boundaries and reduced response in fiber regions. The smoother profiles observed in data can be explained by small mechanical misalignments and unmodeled optical effects, such as wavelength-dependent absorption and realistic surface roughness. Overall, the good qualitative agreement between simulation and experiment validates the implemented optical model and provides a reliable framework for extrapolation to the full-scale SuperFGD.

These results demonstrate the excellent performance of the SuperFGD design in terms of light yield, spatial resolution, and optical isolation. The combined experimental and simulation studies provide a robust foundation for detector calibration, reconstruction algorithms, and modeling in the ND280 upgrade and future high-granularity neutrino detectors.
\label{sec:Conclusion}


\section*{Acknowledgements}

This work is supported in part by the Russian Science Foundation grant number 24-72-10089.





\bibliographystyle{elsarticle-num}
\bibliography{bibliography}

\end{document}